\documentclass[english,pra,two column, single space,floatfix]{revtex4-1}
\usepackage[T1]{fontenc}
\usepackage[latin9]{inputenc}
\usepackage{color}
\usepackage{bm}
\usepackage{amsmath}
\usepackage{amssymb}
\usepackage{graphicx}
\usepackage{esint}

\makeatletter

\newcommand{\lyxdot}{.}

 \@ifundefined{textcolor}{}
 {%
   \definecolor{BLACK}{gray}{0}
   \definecolor{WHITE}{gray}{1}
   \definecolor{RED}{rgb}{1,0,0}
   \definecolor{GREEN}{rgb}{0,1,0}
   \definecolor{BLUE}{rgb}{0,0,1}
   \definecolor{CYAN}{cmyk}{1,0,0,0}
   \definecolor{MAGENTA}{cmyk}{0,1,0,0}
   \definecolor{YELLOW}{cmyk}{0,0,1,0}
 }
\numberwithin{equation}{section}

\makeatother

\usepackage{babel}
\begin{document}

\title{Dynamical preparation of EPR entanglement in two-well Bose-Einstein
condensates}

\author{B. Opanchuk$^{1}$, Q. Y. He$^{1,2}$, M. D. Reid$^{1,*}$, and P.
D. Drummond$^{1,\dagger}$}

\affiliation{$^{1}$Centre for Atom Optics and Ultrafast Spectroscopy, Swinburne
University of Technology, Melbourne 3122, Australia}

\affiliation{$^{2}$State Key Laboratory of Mesoscopic Physics, School of Physics,
Peking University, Beijing 100871, People\textquoteright{}s Republic
of China }

\address{$^{\dagger}$pdrummond@swin.edu.au, $^{*}$mdreid@swin.edu.au}
\begin{abstract}
We propose to generate Einstein-Podolsky-Rosen (EPR) entanglement
between groups of atoms in a two-well Bose-Einstein condensate using
a dynamical process similar to that employed in quantum optics. The
local nonlinear $S$-wave scattering interaction has the effect of
creating a spin squeezing at each well, while the tunneling, analogous
to a beam splitter in optics, introduces an interference between these
fields that results in an inter-well entanglement. We consider two
internal modes at each well, so that the entanglement can be detected
by measuring a reduction in the variances of the sums of local Schwinger
spin observables. As is typical of continuous variable (CV) entanglement,
the entanglement is predicted to increase with atom number, and becomes
sufficiently strong at higher numbers of atoms that the EPR paradox
and steering non-locality can be realized. The entanglement is predicted
using an analytical approach and, for larger atom numbers, stochastic
simulations based on truncated Wigner function. We find generally
that strong tunnelling is favourable, and that entanglement persists
and is even enhanced in the presence of realistic nonlinear losses.
\end{abstract}
\maketitle

\section{Introduction\label{sec:Introduction}}

Entanglement between groups of atoms has been confirmed experimentally,
and recent experiments report the development of quantum correlated
twin-atom beams~\cite{polzik,neweprbec-1,neweprbec2grossent,neweprbecatom entsmerzi}.
This represents a first benchmark for investigations into multi-particle
non-locality that could deepen our understanding of the ``classicality
versus quantumness'' for macroscopic objects~\cite{macroideas}.
So far, however, there has been no reported conclusive demonstration
of stronger forms of quantum non-locality (such as violations of Bell
inequalities~\cite{Bell}, the Einstein-Podolsky-Rosen paradox~\cite{epr,rmp,ou}
or steering~\cite{hw-steering-1,EPRsteering-1,loopholefreesteering})
using mesoscopic groups of atoms, although there have been a number
of theoretical proposals and studies of the correlation between spatially
separated atoms~\cite{GardinerDeuar,Moleccorrel,olsenferris,westbrook,kett4waveexp,bargill,bectheoryepr,coldmolebell}.
Highly efficient detection of an Einstein-Podolsky-Rosen (EPR) paradox
for quadrature field amplitudes~\cite{ou,rmp}, and loophole-free
steering for photons~\cite{loopholefreesteering}, has been been
realized in optics.

In this paper, our goal is to develop a strategy for generating spatial
entanglement between mesoscopic groups of atoms confined to the potential
wells of an ultra-cold Bose Einstein condensate (BEC). Such atoms
have been verified entangled~\cite{esteve}, but we seek to achieve
an unambiguous EPR paradox-steering type of entanglement, in which
the entanglement between the atoms of different wells can be characterized,
and readily extended to situations involving the genuine multi-partite
entanglement of groups of atoms at three or more sites~\cite{multipartent}.
Apart from the potential to test quantum mechanics in new regimes,
this type of entanglement underpins many important applications in
the fields of quantum information and metrology~\cite{eprappli,NoonDowling,steercry,teleeprbeamsplitter}.

There are many possible strategies for the generation of such spatial
multi-atom EPR entanglement. A recent experiment demonstrates EPR-type
correlation near the coherent noise level using spin changing collisions
\cite{neweprbec2grossent}, and there has been a recent proposal to
create spatial entanglement between two wells by direct adiabatic
cooling to the ground state~\cite{ground state}. The most common
method used in quantum optics however is to combine two squeezed single
mode fields through a beam splitter (BS)~\cite{teleeprbeamsplitter,eprbeamsplitter_following,sum,rmp}.
The method relies on a nonlinearity to produce squeezing in each mode
locally, followed by a linear coupling transformation to create the
entanglement between the two modes. 

Motivated by this, we explore in this paper a similar dynamical strategy
applied to the BEC double potential well system. Following Milburn
et al~\cite{milburnmodel}, we assume the atoms of each well can
be modeled using a single mode approximation, and introduce the respective
boson operators, $a$ and $b$. In this case, the S-wave scattering\emph{
}intra-well\emph{ }interactions, given by Hamiltonians $\hat{H}=\hbar g\hat{a}^{\dagger2}\hat{a}^{2}$
and $\hat{H}=\hbar g\hat{b}^{\dagger2}\hat{b}^{2}$, provide the nonlinearity
at each well that generates a local spin squeezing, while the coupling
or tunneling inter-well term, modeled as $\hat{H}=\hbar\kappa(\hat{a}^{\dagger}\hat{b}+\hat{a}\hat{b}^{\dagger})$,
provides the linear beam-splitter transformation~\cite{bs,wellbecmurr}
that generates inter-well two-mode entanglement. 

We consider in fact two dynamical strategies: in the first the local
nonlinear and nonlocal tunnelling processes act sequentially\emph{;}
in the second they act simultaneously. While the second strategy is
likely to be more practical, we analyze the sequential case first,
in the earlier sections of the paper, because it allows a full quantum
solution within the constraints of the two-mode model.

We show that substantial entanglement can be generated in both cases,
provided parameters are optimized. An analysis of what is accessible
experimentally indicates that this entanglement could be within reach
of current set-ups, though the realization of EPR and steering paradoxes
presents a greater challenge, and may require more sophisticated experimental
procedures. 

Perhaps surprisingly, we find that a large amount of entanglement
requires using large numbers of atoms. This result is verified using
both a full quantum analysis and the truncated Wigner function~\cite{wigner}
which becomes valid as $1/N^{3/2}\rightarrow0$. The manifestation
of a continuous variable (CV) EPR paradox has been shown for large
optical amplitudes~\cite{RDeprlarge}, but may have been thought
impossible to realize for large groups of atoms. The prediction is
consistent with those of Ferris et al~\cite{olsenferris} and who
study EPR entanglement in BEC four wave mixing, though in our case
an atomic homodyne detection~\cite{neweprbec2grossent} is not used,
the EPR observables being the local Schwinger spins that can be measured
via Rabi rotation and atom counting. 

The paper is organized as follows. We first summarize in Section II
the meaning of Einstein-Podolsky-Rosen entanglement, and outline how
it is to be detected. The dynamical solutions for the entanglement
via the two-step method are presented in Sections~\ref{sec:Dynamical-preparation-of}
and~\ref{sec:Two-step-dynamical-entanglement}. The truncated Wigner
function simulations and the results for the inter-well entanglement
via simultaneous evolution are treated in Section~\ref{sec:Truncated-Wigner-application}.
An analysis modeling current experimental regimes, including the effect
of nonlinear losses, indicates that the EPR entanglement is robust
against the expected decoherence effects.

\section{EPR entanglement\label{sec:Measurement-strategies}}

The original CV EPR paradox~\cite{epr} considers correlations between
the positions and momenta of two particles emitted from a source.
With optical or atomic Bose fields, one can define the quadrature
phase amplitudes of two spatially separated field modes, as $\hat{X}_{A}=(\hat{a}^{\dagger}+\hat{a})/2$,
and $\hat{P}_{A}=(\hat{a}^{\dagger}-\hat{a})/2i$, and $\hat{X}_{B}=(\hat{b}^{\dagger}+\hat{b})/2$,
and $\hat{P}_{B}=(\hat{b}^{\dagger}-\hat{b})/2i$. These are analogous
to the position and momentum in the two-particle system. 

The EPR paradox arises when both $\hat{X}_{A}$ and $\hat{X}_{B}$,
and $\hat{P}_{A}$ and $\hat{P}_{B}$, are maximally correlated, so
that measurement of $\hat{X}_{A}$ enables exact prediction of\emph{
}the outcome for measurement of $\hat{X}_{B}$, and the measurement
of $\hat{P}_{A}$ enables exact prediction of the outcome of measurement
of $\hat{P}_{B}$. EPR argued that, assuming no ``spooky action at
distance'', the action of measuring $\hat{X}_{A}$ could not ``create''
the result for $\hat{X}_{B}$. They then concluded, since the result
for $\hat{X}_{B}$ can be predicted without disturbance of that system,
that the result for the outcome is predetermined. Since the predetermination
of both $\hat{X}_{B}$ and $\hat{P}_{B}$ is without uncertainty,
there can be no equivalent local quantum state interpretation. Hence,
EPR argued that quantum mechanics was incomplete. The premises assumed
by EPR are often termed ``local realism'', and the EPR paradox can
be thought of as a demonstration of the incompatibility between ``local
realism'' and the ``completeness of quantum mechanics''.

EPR's argument applies when one observer (Alice) can make precise
predictions for the outcome of measurements made by a second, distant
observer (Bob). The key issue for the EPR paradox is that Alice can
infer a result for either of two of Bob's conjugate observables, by
measuring locally on her system. The EPR paradox arises when the accuracy
of her inferences would violate quantum mechanics, if she could infer
results for both conjugate observables simultaneously. This demonstration
of the EPR paradox is most simply achieved by comparing the conditional
variances for Alice's measurements with the variances of the Heisenberg
uncertainty relation~\cite{rmp,eprcrit}.

\subsection{Entanglement criteria\emph{ }}

The original EPR paradox focused on states that showed correlation
and anti-correlation for position and momentum respectively. Duan
et al and Simon~\cite{simon-1,Duan-simon-1} showed that entanglement
between modes $a$ and $b$ is confirmed if 
\begin{equation}
D=\Delta^{2}(\hat{X}_{A}-\hat{X}_{B})+\Delta^{2}(\hat{P}_{A}+\hat{P}_{B})<1.\label{eq:duan-1}
\end{equation}
In the case~(\ref{eq:duan-1}), the $1$ arises from the commutation
relation $[\hat{a},\hat{a}^{\dagger}]=1$, and reflects the quantum
noise associated with the four observables. A more general criterion
is the product form: entanglement is confirmed when $\Delta(\hat{X}_{A}-\hat{X}_{B})\Delta(\hat{P}_{A}+\hat{P}_{B})<1/2$~\cite{proof for product form,sum}. 

Quadrature phase amplitudes are measured using local oscillator methods,
in which a strong field interferes with a signal field, using a beam
splitter. A full analysis of a local oscillator measurement shows
that it is actually equivalent to a Schwinger spin measurement, once
the local oscillator is accounted for. It is useful to consider entanglement
measures that have been developed for spin measurements. In particular,
one can show entanglement using the spin version of~(\ref{eq:duan-1})~\cite{sum},
\begin{equation}
\Delta^{2}(\hat{J}_{A}^{X}\mp\hat{J}_{B}^{X})+\Delta^{2}(\hat{J}_{A}^{Y}\pm\hat{J}_{B}^{Y})<|\langle\hat{J}_{A}^{Z}\rangle|+|\langle\hat{J}_{B}^{Z}\rangle|,\label{eq:duan-3-1-1}
\end{equation}
and also the Heisenberg-product entanglement criterion~\cite{proof for product form}
\begin{equation}
\sqrt{\Delta^{2}(\hat{J}_{A}^{\theta}-\hat{J}_{B}^{\theta})\cdot\Delta^{2}(\hat{J}_{A}^{\theta+\pi/2}+\hat{J}_{B}^{\theta+\pi/2})}<\frac{|\langle\hat{J}_{A}^{Y}\rangle|+|\langle\hat{J}_{B}^{Y}\rangle|}{2}.\label{eq:product form-1}
\end{equation}

\subsection{EPR paradox steering criteria\emph{ }}

While entanglement as confirmed by~(\ref{eq:duan-1}) is necessary
for the EPR paradox, it is not sufficient. To quantitatively demonstrate
the paradox, in the style constructed by EPR, the level of correlation
in Alice's predictions must be compared with the quantum limit for
a local state that might predetermine Bob's statistics. Thus, for
the EPR paradox, the relevant quantum noise level is that of one observer,
$B$, alone. 

The EPR paradox confirms ``steering'', whereas the entanglement
of~(\ref{eq:duan-1}) does not. Steering has been established as
a distinct form of non-locality~\cite{hw-steering-1}. The EPR paradox
and steering types of entanglement provide a distinct resource for
quantum information with applications not achievable for arbitrary
entangled states~\cite{steercry}.

An EPR paradox signature has been formulated in terms of conditional
variances~\cite{eprcrit}: thus EPR entanglement is observed when
\begin{equation}
\Delta(\hat{X}_{B}|\hat{X}_{A})\Delta(\hat{P}_{B}|\hat{P}_{A})<1/4.\label{eq:eprcritpordxp}
\end{equation}
Since the precise choice of measurement at mode $A$ is not important,
only the inference, this criterion is sometimes more generally written
as $\Delta_{\mathrm{inf}}\hat{X}_{B}\Delta_{\mathrm{inf}}\hat{P}_{B}<1$. 

More recently, it has been pointed out that the EPR signature is achieved
once the entanglement variances become small enough~\cite{olsenferris,rmp}.
EPR entanglement is detected, if either 
\begin{equation}
D=\Delta^{2}(\hat{X}_{A}-\hat{X}_{B})+\Delta^{2}(\hat{P}_{A}+\hat{P}_{B})<1/2,\label{eq:eprhalfduan}
\end{equation}
or $\Delta(\hat{X}_{A}-\hat{X}_{B})\Delta(\hat{P}_{A}+\hat{P}_{B})<1/4$.
These criteria are special cases of~(\ref{eq:eprcritpordxp}), and
may not give the optimal measurement strategy for obtaining an EPR
paradox, but are useful in many practical cases where only~(\ref{eq:eprhalfduan})
is measured.

Spin EPR measures can also be obtained, just as with entanglement
\emph{per se}, using the inferred Heisenberg uncertainty principle
approach. From the Heisenberg spin uncertainty relation $\Delta\hat{J}_{B}^{X}\Delta\hat{J}_{B}^{Y}\geq|\langle\hat{J}_{B}^{Z}\rangle|/2$,
one can derive several simple spin-EPR criteria~\cite{rmp,sum,caval reid uncerspin}.
For the case of a large non-zero mean spin, taken for definiteness
in the $z-$direction, an EPR paradox is demonstrated when 
\begin{equation}
\Delta_{\mathrm{inf}}\hat{J}_{B}^{X}\Delta_{\mathrm{inf}}\hat{J}_{B}^{Y}<\frac{1}{2}|\langle\hat{J}_{B}^{Z}\rangle|.\label{eq:spinepr}
\end{equation}
This criterion for the EPR paradox can be expressed in a particularly
useful, though less general form, as follows:
\begin{equation}
\sqrt{\Delta^{2}(\hat{J}_{A}^{\theta}-g\hat{J}_{B}^{\theta})\cdot\Delta^{2}(\hat{J}_{A}^{\theta+\pi/2}+g^{\prime}\hat{J}_{B}^{\theta+\pi/2})}<\frac{|\langle\hat{J}_{B}^{Y}\rangle|}{2}.\label{eq:product form epr-1}
\end{equation}
In this version, the simplification has been made that the inference
is given using a fixed linear approximation with constant gain $g$.

\section{Dynamical preparation of BEC EPR entanglement\label{sec:Dynamical-preparation-of}}

We now turn to examining how to generate EPR entanglement in a BEC.
A two-well BEC system can be modeled by the two-mode Hamiltonian~\cite{milburnmodel}:
\begin{equation}
\hat{H}/\hbar=\kappa\left(\hat{a}^{\dagger}\hat{b}+\hat{a}\hat{b}^{\dagger}\right)+\frac{g}{2}\left(\hat{a}^{\dagger}\hat{a}^{\dagger}\hat{a}\hat{a}+\hat{b}^{\dagger}\hat{b}^{\dagger}\hat{b}\hat{b}\right).\label{hamgs-1-1}
\end{equation}
Here $\kappa$ is the conversion rate between the two components,
denoted by mode operators $a$ and $b$, and $g$ is the nonlinear
self interaction coefficient proportional to the three-dimensional
S-wave scattering length, $a_{3D}$. The first term proportional to
$\kappa$ describes an exchange of particles between the two wells
(modes) in which total number is conserved. This linear term is equivalent
to that for an optical beam splitter~\cite{bs,wellbecmurr}. The
Hamiltonian model applies to other two-mode bosonic systems including
optical cavity modes or superconducting wave-guides with a nonlinear
medium, as well as other types of two-mode atomic systems. However,
we will use a two-well picture to illustrate the gedanken-experiment
we have in mind, even though implementations may not be exactly in
this form. The two-well model and system has been studied extensively
in relation to macroscopic superposition states~\cite{bec well ham cirac,wellbecmurr}
and ultra-sensitive interferometric measurement~\cite{Leeprl ,Gross2010}.

\begin{figure}
\begin{centering}
\includegraphics[width=0.4\columnwidth]{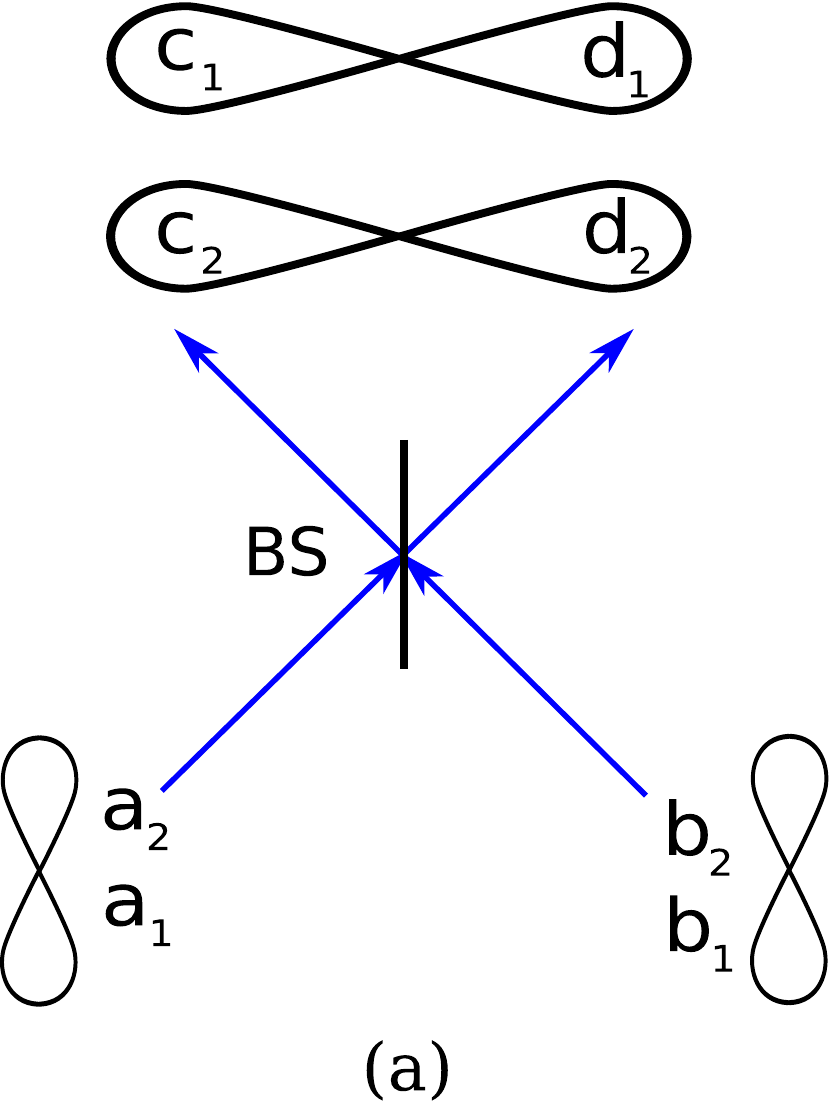}\quad{}\includegraphics[width=0.4\columnwidth]{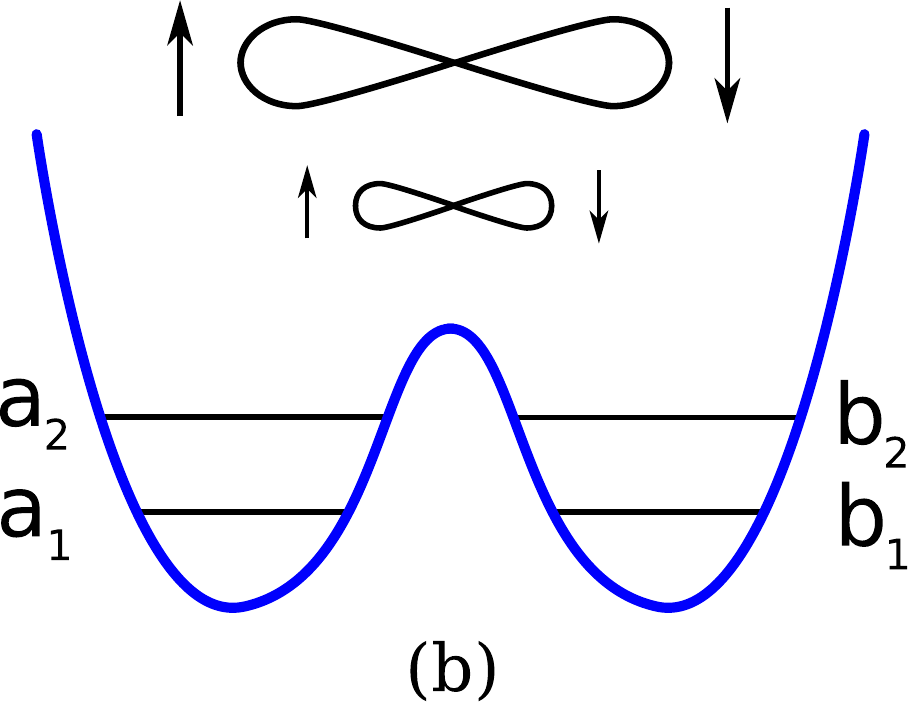}
\par\end{centering}

\caption{Creation of spatial entanglement between modes $a_{i}$ and $b_{i}$
at different sites. The optical scheme is depicted in (a) and the
BEC two-well proposal in (b). Local spin squeezing can be produced
when the two modes at each well evolve for a time $\tau$ under the
nonlinear Hamiltonian~(\ref{eq:ham1}). Depending on the exact configuration,
there can be coupling terms between the two modes at each site. The
well height is controlled to allow a tunneling cross-interaction between
the modes. This is equivalent to the modes interfering using a beam
splitter, and the effect is to generate an inter-well spatial entanglement.~\label{fig:Entanglement-generation-by-BS}}
\end{figure}

\textcolor{black}{If, in addition, we allow two internal spin components
per well, the Hamiltonian becomes:
\begin{equation}
\hat{H}/\hbar=\sum_{i}\kappa_{i}\hat{a}_{i}^{\dagger}\hat{b}_{i}+\frac{1}{2}\sum_{ij}g_{ij}\hat{a}_{i}^{\dagger}\hat{a}_{j}^{\dagger}\hat{a}_{j}\hat{a}_{i}+\left\{ \hat{a}_{i}\leftrightarrow\hat{b}_{i}\right\} .\label{eq:Hamiltonian}
\end{equation}
Here we consider two internal modes at each EPR site $A$ and $B$,
with four modes in total, as shown schematically in Fig.~\ref{fig:Entanglement-generation-by-BS}.
The local modes can be independent (in which case local cross couplings
$g_{ij}$ are zero), or not independent, as where the modes are coupled
by the BEC self interaction term}~\cite{Gross2010}.\textcolor{black}{{}
The coupling constant is proportional to the three-dimensional S-wave
scattering length, so that $g_{ij}\propto a_{ij}$, as in the two-mode
case. }

\textcolor{black}{We will describe our results using} the equivalent
dimensionless Hamiltonian 
\begin{equation}
\tilde{H}=\sum_{i}\tilde{\kappa}_{i}\hat{a}_{i}^{\dagger}\hat{b}_{i}+\frac{1}{2}\sum_{ij}\tilde{g}_{ij}\hat{a}_{i}^{\dagger}\hat{a}_{j}^{\dagger}\hat{a}_{j}\hat{a}_{i}+\left\{ \hat{a}_{i}\leftrightarrow\hat{b}_{i}\right\} ,\label{eq:dimensionless-H}
\end{equation}
with dimensionless coefficients $\tilde{\kappa}_{i}=\kappa_{i}/g_{11}N_{A}$,
$\tilde{g}_{ij}=g_{ij}/g_{11}N_{A}$, where $N_{A}$ is the initial
total boson number in well A, and a corresponding dimensionless time
$\tau=g_{11}N_{A}t$.\textcolor{black}{{}  For definiteness, we will
choose ratios of nonlinear couplings $\tilde{g}_{ij}$ to correspond
to known S-wave scattering lengths of $^{87}$Rb, (between }$|1\rangle\equiv|F=1,\, m_{F}=+1\rangle$
and $|2\rangle\equiv|F=2,\, m_{F}=-1\rangle$)\textcolor{black}{{} at
specific magnetic field strengths near a Feshbach resonance $B=9.105\,\mathrm{G}$.
Namely, we use $B=9.086\,\mathrm{G}$ with corresponding $a_{12}=107.8a_{0}$,
and $B=9.116\,\mathrm{G}$ with corresponding $a_{12}=80.8a_{0}$}~\cite{kaufman}\textcolor{black}{,
where $a_{0}$ is the Bohr radius. Intra-species scattering length
are constant and equal to $a_{11}=100.4a_{0}$ and $a_{22}=95.5a_{0}$.
Other results will be given as well.}

\textcolor{black}{We propose to generate EPR states by preparing the
system in a multi-mode coherent state and then allowing evolution
according to the Hamiltonian (\ref{hamgs-1-1}). Two strategies are
compared.}\textcolor{red}{{} }\textcolor{black}{Firstly, where a manipulation
of tunneling (e.g.},\textcolor{black}{{} by changing the potential barrier
height) occurs, so that the nonlinear and linear terms are applied
sequentially, and secondly, where the nonlinear and linear tunneling
terms act simultaneously, as in~(\ref{hamgs-1-1}\textendash{}\ref{eq:dimensionless-H}).
The second strategy is likely to be more readily implemented, and
is solved for in Section~\ref{sec:Truncated-Wigner-application}
via a truncated Wigner function method, with the inclusion of further
effects such as nonlinear losses.}

\textcolor{black}{In the two-step strategy, local squeezing is first
generated for each mode $a$ and $b$ via the nonlinear Hamiltonian
in the absence of tunneling, and subsequently, a strong tunneling
interaction provides a linear mixing that generates an EPR entanglement
between the two new modes $c$ and $d$. The strategy is depicted
in Fig.~\ref{fig:Entanglement-generation-by-BS}. }The technique
is similar to that investigated experimentally in fibre-optics entanglement~\cite{fiber experiment}.
An important difference is that the fiber experiment used time-delayed
pulses and dispersion to eliminate interactions, that is local cross
couplings, between the components. This is not readily feasible in
BEC experiments, although Feshbach resonances can achieve this to
a limited extent. In the next Section, we present full quantum solutions
for this two-step strategy.

We briefly remark on our choice of initial conditions. In experiments
with multiple wells, it is generally possible to prepare condensates
with relative phase coherence, provided that tunneling is strong enough
during the evaporative cooling process. The overall phase is random,
with a total number uncertainty that is typically at least Poissonian.
This quantum state is therefore well represented by a mixture of coherent
states with a random overll phase. However, since none of our results
depend on the overall phase, it is sufficient to consider just a single
overall coherent state with an arbitrary phase. This is a low-temperature
limit, which will develop additional fluctuations as temperatures
are increased to the critical temperature.

\section{Strategy I: Two-step dynamical entanglement generation\label{sec:Two-step-dynamical-entanglement}}

\subsection{Generation of local spin squeezing}

In the first step, squeezing is generated locally via a nonlinearity,
given by the Hamiltonian
\begin{equation}
\tilde{\mathcal{H}}=\sum_{i,j}\frac{\tilde{g}_{ij}}{2}\hat{a}_{i}^{\dagger}\hat{a}_{j}^{\dagger}\hat{a}_{j}\hat{a}_{i}.\label{eq:ham1}
\end{equation}
The initial state is a product coherent state for each mode: $|\alpha/\sqrt{2}\rangle_{a_{1}}|\alpha/\sqrt{2}\rangle_{a_{2}}$.
This models the relative coherence between the wells obtained with
a low inter-well potential barrier, together with an overall Poissonian
number fluctuation that is typically found in an experimental BEC.
We explain in the Appendix how to calculate the nonlinear quantum
dynamical solutions of~(\ref{eq:ham1}). These are exact calculations,
provided the original multimode Hamiltonian can be reduced to simple
one or two-mode forms.

Let $\hat{a}_{1},$$\hat{a}_{2}$ be operators for the two internal
states at well $A$, and $\hat{b}_{1}$, $\hat{b}_{2}$ operators
for two internal states at well $B$. Here $\hat{N}_{A}=\hat{a}_{2}^{\dagger}\hat{a}_{2}+\hat{a}_{1}^{\dagger}\hat{a}_{1}$
and $\hat{N}_{B}=\hat{b}_{2}^{\dagger}\hat{b}_{2}+\hat{b}_{1}^{\dagger}\hat{b}_{1}$
are the atom number operators of these modes in each well. We introduce
a phase-rotated Schwinger spin operator measurement that can be performed
at each site. For site $A$, we define
\begin{equation}
\begin{split}\hat{J}_{A}^{X} & =\frac{1}{2}\left(\hat{a}_{2}^{\dagger}\hat{a}_{1}e^{i\Delta\theta}+\hat{a}_{1}^{\dagger}\hat{a}_{2}e^{-i\Delta\theta}\right),\\
\hat{J}_{A}^{Y} & =\frac{1}{2i}\left(\hat{a}_{2}^{\dagger}\hat{a}_{1}e^{i\Delta\theta}-\hat{a}_{1}^{\dagger}\hat{a}_{2}e^{-i\Delta\theta}\right),\\
\hat{J}_{A}^{Z} & =\frac{1}{2}\left(\hat{a}_{2}^{\dagger}\hat{a}_{2}-\hat{a}_{1}^{\dagger}\hat{a}_{1}\right),
\end{split}
\label{eq:spinsc}
\end{equation}
where $\Delta\theta=\theta_{2}-\theta_{1}$ is the phase shift between
the mode $1$ and mode $2$. There is also an analogous definition
at $B$ with phase shift $\Delta\phi=\phi_{2}-\phi_{1}$.

We select the phase shift to ensure $\langle\hat{J}^{Y}\rangle\neq0$,
and the Schwinger spin operators orthogonal to $\hat{J}^{Y}$ are
given by
\begin{equation}
\hat{J}^{\theta}=\cos(\theta)\hat{J}^{Z}+\sin(\theta)\hat{J}^{X}\label{eq:spinrot}
\end{equation}
all of which have the property $\langle\hat{J}^{\theta}\rangle=0$.
Here $\Delta\theta=\Delta\phi=\pi/2-\alpha$, and $\alpha$ is time
dependent given by the character of $\langle\hat{a}_{2}^{\dagger}\hat{a}_{1}\rangle=|\langle\hat{a}_{2}^{\dagger}\hat{a}_{1}\rangle|e^{i\alpha}$.
This plane contains an infinite family of maximally conjugate Schwinger
spin operators, generally given by $\hat{J}^{\theta}$ and $\hat{J}^{\theta+\pi/2}$
which obey the Heisenberg uncertainty relation
\begin{equation}
\Delta\hat{J}^{\theta}\Delta\hat{J}^{(\theta+\pi/2)}\geq|\langle\hat{J}^{Y}\rangle|/2.\label{eq:spinuncer}
\end{equation}
Quantum squeezing occurs when the variance in one conjugate observable
is reduced below the Heisenberg limit. Thus,
\begin{equation}
\Delta^{2}\hat{J}^{\theta}<|\langle\hat{J}^{Y}\rangle|/2\label{eq:spinsqu}
\end{equation}
is said to be a ``spin squeezed'' state~\cite{uedakitspinsq,wineland e,Spin Squeezing}. 

\begin{figure}[h]
\begin{centering}
\includegraphics[width=1\columnwidth]{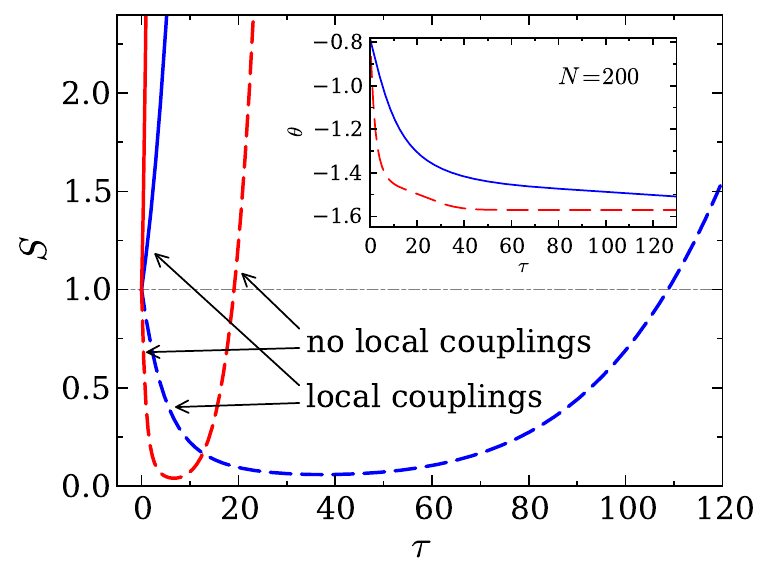}
\par\end{centering}

\centering{}\caption{(Color online) Squeezing of local Schwinger spin operators versus
interaction time $\tau$. The plot shows the squeezing $\Delta^{2}\hat{J}^{\theta}/n_{0}$
(solid lines), and $\Delta^{2}\hat{J}^{\theta+\pi/2}/n_{0}$ (dashed
lines), where the shot noise level is $n_{0}=|\langle\hat{J}^{Y}\rangle|/2$.
Squeezing is obtained when $S=\Delta^{2}\hat{J}^{\theta,\ \theta+\pi/2}/(|\langle\hat{J}^{Y}\rangle|/2)<1$.
Inset shows the optimal phase choice $\theta$ for squeezing. The
dimensionless coupling parameters correspond to $^{87}$Rb atoms at
magnetic field $B=9.116\,\mathrm{G}$ (blue lines), with corresponding
scattering lengths, as explained in the text. We also give results
for the case without local cross couplings (red lines), i.e., $g_{12}=0,\ g_{22}=g_{11}$.
Here $N_{A}=200$. \label{fig:squeezing-after-BEC.} }
\end{figure}

Figure~\ref{fig:squeezing-after-BEC.} shows the prediction for dynamical
spin-squeezing, according to the solutions explained in the Appendix
A, based on the Hamiltonian Eq.~(\ref{eq:ham1}). Here we have assumed
that $a_{1},b_{1}$ and $a_{2},b_{2}$ are initially in coherent states.
For simplicity, we suppose that the initial state is prepared in a
four-mode coherent state by using a Rabi rotation: $|\psi\rangle=|\frac{\alpha}{\sqrt{2}}\rangle_{a_{1}}|\frac{\alpha}{\sqrt{2}}\rangle_{a_{2}}|\frac{\alpha}{\sqrt{2}}\rangle_{b_{1}}|\frac{\alpha}{\sqrt{2}}\rangle_{b_{2}}$.
After preparation, we assume that the inter-well potential is increased
so that each well evolves independently, to give the solutions. 

We have considered the conditions required to obtain the best squeezing
of Schwinger spin operators by optimizing the phase choice $\theta$
(inset of Fig.~\ref{fig:squeezing-after-BEC.}). We set
\begin{eqnarray}
\frac{\partial\Delta^{2}\hat{J}^{\theta}}{\partial\theta} & = & \frac{\partial}{\partial\theta}[\cos^{2}\theta\Delta^{2}\hat{J}^{Z}+\sin^{2}\theta\Delta^{2}\hat{J}^{X}\nonumber \\
 &  & +2\cos\theta\sin\theta\langle\hat{J}^{Z},\ \hat{J}^{X}\rangle]\nonumber \\
 & = & 2\cos(2\theta)\langle\hat{J}^{Z},\ \hat{J}^{X}\rangle-\sin(2\theta)(\Delta^{2}\hat{J}^{Z}-\Delta^{2}\hat{J}^{X})\nonumber \\
 & = & 0\label{eq:1}
\end{eqnarray}
and therefore obtain as the optimal squeezing angle
\begin{equation}
\theta=\frac{1}{2}\arctan\left(\frac{2\langle\hat{J}_{A}^{Z},\hat{J}_{A}^{X}\rangle}{\Delta^{2}\hat{J}_{A}^{Z}-\Delta^{2}\hat{J}_{A}^{X}}\right),\label{eqn:single-well:angle-1-1}
\end{equation}
where
\begin{equation}
\langle\hat{J}_{A}^{Z},\hat{J}_{A}^{X}\rangle=\frac{1}{2}\left(\langle\hat{J}_{A}^{Z}\hat{J}_{A}^{X}\rangle+\langle\hat{J}_{A}^{X}\hat{J}_{A}^{Z}\rangle-2\langle\hat{J}_{A}^{Z}\rangle\langle\hat{J}_{A}^{X}\rangle\right).
\end{equation}
The squeezing value in this case is:
\begin{equation}
S^{\theta,\theta+\pi/2}=\frac{\Delta^{2}\hat{J}_{A}^{\theta,\theta+\pi/2}}{\vert\langle\hat{J}_{A}^{Y}\rangle\vert/2},\label{eqn:single-well:squeezing-1-1}
\end{equation}
where
\begin{equation}
\Delta^{2}\hat{J}_{A}^{\theta}=\cos^{2}\theta\Delta^{2}\hat{J}_{A}^{Z}+\sin^{2}\theta\Delta^{2}\hat{J}_{A}^{X}+2\sin\theta\cos\theta\langle\hat{J}_{A}^{Z},\hat{J}_{A}^{X}\rangle.
\end{equation}

\subsection{Producing the spatial entanglement}

The next step, after generating the local squeezing at each well,
is to decrease the inter-well potential for a short time, so that
it acts as a controllable, non-adiabatic beam splitter~\cite{wellbecmurr},
to allow interference between the wells; other methods of obtaining
an effective beam-splitter could also be feasible. 

Entanglement is generated by the interference of two squeezed states
on a $50:50$ beamsplitter with a relative optical phase of $\varphi$
(Fig.~\ref{fig:Entanglement-generation-by-BS}). The resulting entangled
modes are labelled by $c_{1,2}$ and $d_{1,2}$. Schwinger spin operators
$\hat{J}_{C/D}^{\theta}$ are defined for these modes, in accordance
with~Eq. (\ref{eq:spinsc}). We note that $ $$\hat{J}_{C}^{\theta,\theta+\pi/2}$,
$\hat{J}_{D}^{\theta,\theta+\pi/2}$ are measurable locally, in the
style necessary for an EPR experiment, by using Rabi rotations and
number measurements.

The input-output relations for the Schwinger spin operators are given
by $\hat{c}=t\hat{a}+re^{i\varphi}\hat{b}$, and $\hat{d}=t\hat{b}-re^{-i\varphi}\hat{a}$
with the amplitude reflection and transmission coefficients being
denoted $t$ and $r$. Here we use $\varphi=\pi/2$ and $r=t=1/\sqrt{2}$.
Thus:
\begin{equation}
\hat{c}=\frac{\hat{a}+i\hat{b}}{\sqrt{2}},\,\hat{d}=\frac{\hat{b}+i\hat{a}}{\sqrt{2}}.\label{eq:bms}
\end{equation}

The variances $\Delta^{2}(\hat{J}_{C}^{\theta}-\hat{J}_{D}^{\theta})$
and $\Delta^{2}(\hat{J}_{C}^{\theta+\pi/2}+\hat{J}_{D}^{\theta+\pi/2})$
can \emph{both} be small, so that 
\begin{eqnarray}
S_{-}=\Delta^{2}(\hat{J}_{C}^{\theta}-\hat{J}_{D}^{\theta}) & < & \frac{1}{2}(|\langle\hat{J}_{C}^{Y}\rangle|+|\langle\hat{J}_{D}^{Y}\rangle|)
\end{eqnarray}
 and 
\begin{eqnarray*}
S_{+} & = & \Delta^{2}(\hat{J}_{C}^{\theta+\pi/2}+\hat{J}_{D}^{\theta+\pi/2})<\frac{1}{2}(|\langle\hat{J}_{C}^{Y}\rangle|+|\langle\hat{J}_{D}^{Y}\rangle|),
\end{eqnarray*}
the degree of variance reduction being limited only by the amount
of squeezing in the input modes. This is the signature of EPR entanglement,
in accordance with the entanglement criteria~(\ref{eq:duan-3-1-1})
and~(\ref{eq:product form-1}). In fact,
\begin{eqnarray}
\Delta^{2}(\hat{J}_{C}^{\theta}\mp\hat{J}_{D}^{\theta}) & = & \cos^{2}\theta\Delta^{2}(\hat{J}_{C}^{Z}\mp\hat{J}_{D}^{Z})+\sin^{2}\theta\Delta^{2}(\hat{J}_{C}^{X}\mp\hat{J}_{D}^{X})\nonumber \\
 &  & +\cos\theta\sin\theta[\langle\hat{J}_{C}^{Z}\mp\hat{J}_{D}^{Z},\hat{J}_{C}^{X}\mp\hat{J}_{D}^{X}\rangle\nonumber \\
 &  & +\langle\hat{J}_{C}^{X}\mp\hat{J}_{D}^{X},\hat{J}_{C}^{Z}\mp\hat{J}_{D}^{Z}\rangle].
\end{eqnarray}
A similar expression can be given for $\Delta^{2}(\hat{J}_{C}^{\theta'}\pm\hat{J}_{D}^{\theta'})$.
The solutions show a reduction in these variances, for suitable parameters,
to indicate that Alice can infer Bob's observable $\hat{J}_{C}^{\theta}$
(by measuring $\hat{J}_{D}^{\theta}$) with increasing accuracy, as
$\Delta^{2}(\hat{J}_{C}^{\theta}-\hat{J}_{D}^{\theta})\rightarrow0$.
Similarly, she can infer the conjugate observable $\hat{J}_{C}^{\theta+\pi/2}$
(by measuring $\hat{J}_{D}^{\theta+\pi/2}$) to an increasing accuracy,
as $\Delta^{2}(\hat{J}_{C}^{\theta+\pi/2}+\hat{J}_{D}^{\theta+\pi/2})\rightarrow0$.
We call these variances ``inference variances'' to remind us of
their role in the EPR paradox. Ideally, we want to find a regime for
which both inference variances become very small.

\begin{figure}[t]
\begin{centering}
\includegraphics[width=1\columnwidth]{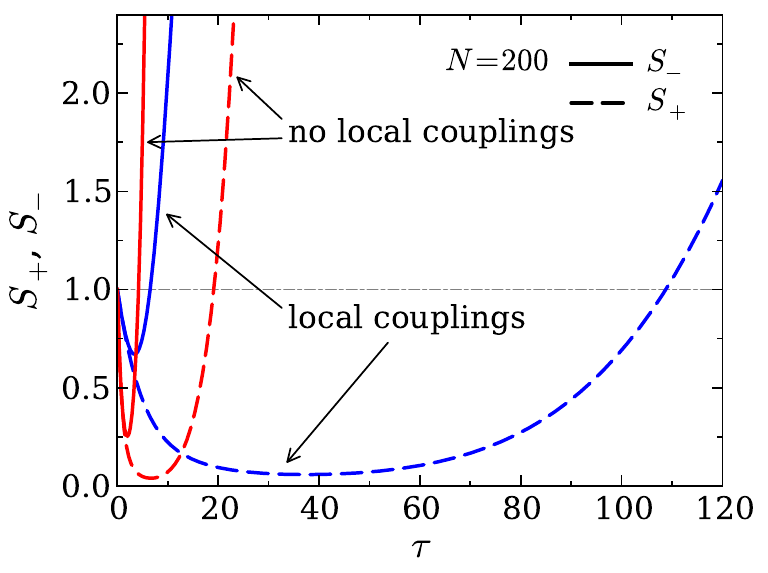}
\par\end{centering}

\centering{}\caption{(Color online) Inference squeezing: $S_{-}=\left[\Delta^{2}(\hat{J}_{C}^{\theta}-\hat{J}_{D}^{\theta})/n_{0}\right]$
(solid lines) and $S_{+}=\left[\Delta^{2}(\hat{J}_{C}^{\theta+\pi/2}+\hat{J}_{D}^{\theta+\pi/2})/n_{0}\right]$
(dashed lines) after the beamsplitter interaction, where the shot
noise level is $n_{0}=(|\langle\hat{J}_{C}^{Y}\rangle|+|\langle\hat{J}_{D}^{Y}\rangle|)/2$.
Here $N=N_{A}=N_{B}=200$. The parameters are the same as Fig.~\ref{fig:squeezing-after-BEC.}.
The red lines show the result without local cross couplings, i.e.,
$g_{12}=0,\ g_{22}=g_{11}$.\label{fig:Inference-squeezing-()}}
\end{figure}

It is convenient to express the output spin operators in terms of
the inputs: we find
\begin{eqnarray}
\hat{J}_{C}^{Z}-\hat{J}_{D}^{Z} & = & \frac{i}{2}[\hat{a}_{2}^{\dagger}\hat{b}_{2}-\hat{b}_{2}^{\dagger}\hat{a}_{2}-\hat{a}_{1}^{\dagger}\hat{b}_{1}+\hat{b}_{1}^{\dagger}\hat{a}_{1}],\nonumber \\
\hat{J}_{C}^{Z}+\hat{J}_{D}^{Z} & = & \frac{1}{2}[\hat{a}_{2}^{\dagger}\hat{a}_{2}-\hat{a}_{1}^{\dagger}\hat{a}_{1}+\hat{b}_{2}^{\dagger}\hat{b}_{2}-\hat{b}_{1}^{\dagger}\hat{b}_{1}],\nonumber \\
\hat{J}_{C}^{X}-\hat{J}_{D}^{X} & = & \frac{i}{2}[e^{i(\theta_{2}-\phi_{1})}\hat{a}_{2}^{\dagger}\hat{b}_{1}-e^{-i(\theta_{1}-\phi_{2})}\hat{b}_{2}^{\dagger}\hat{a}_{1}\nonumber \\
 &  & -e^{i(\theta_{1}-\phi_{2})}\hat{a}_{1}^{\dagger}\hat{b}_{2}+e^{-i(\theta_{2}-\phi_{1})}\hat{b}_{1}^{\dagger}\hat{a}_{2}],\nonumber \\
\hat{J}_{A}^{Z}+\hat{J}_{B}^{Z} & = & \hat{J}_{C}^{Z}+\hat{J}_{D}^{Z},
\end{eqnarray}
where we have used $\theta_{2}-\phi_{2}=\theta_{1}-\phi_{1}=0$ due
to the symmetry of $a$ and $b$, and $\theta_{2}-\phi_{1}=\theta_{2}-\theta_{1}=\pi/2-\alpha$
as introduced for Eq.~(\ref{eq:spinrot}). The solutions for the
inference squeezing $\Delta^{2}(\hat{J}_{C}^{\theta}-\hat{J}_{D}^{\theta})$
and $\Delta^{2}(\hat{J}_{C}^{\theta+\pi/2}+\hat{J}_{D}^{\theta+\pi/2})$
are shown in Fig.~\ref{fig:Inference-squeezing-()}. We note that
unlike for most EPR states, the inference variances are asymmetrical.

The spin orientation measured at each site can be selected independently
to optimize the criterion for the state used. One can then show EPR
entanglement via spin measurements by using the product entanglement
criterion Eq. (\ref{eq:product form-1}):
\begin{equation}
E_{\mathrm{product}}=\frac{\sqrt{\Delta^{2}(\hat{J}_{C}^{\theta}-\hat{J}_{D}^{\theta})\cdot\Delta^{2}(\hat{J}_{C}^{\theta+\pi/2}+\hat{J}_{D}^{\theta+\pi/2})}}{\left(|\langle\hat{J}_{C}^{Y}\rangle|+|\langle\hat{J}_{D}^{Y}\rangle|\right)/2}<1.\label{eq:product form}
\end{equation}

\begin{figure}[t]
\begin{centering}
\includegraphics[width=1\columnwidth]{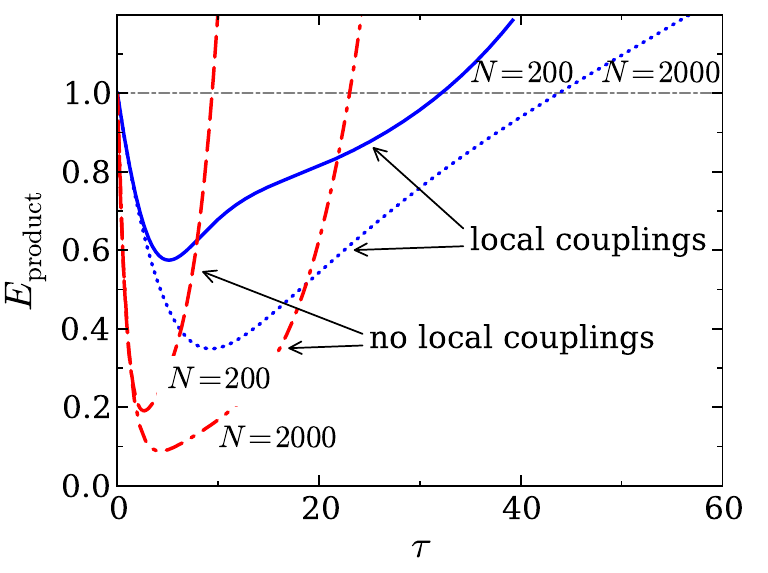}
\par\end{centering}

\caption{Entanglement ($E_{\mathrm{product}}<1$) based on the criterion in
product form~(\ref{eq:product form}). EPR paradox entanglement is
obtained when $E_{\mathrm{product}}<0.5$. \textcolor{black}{The solid
and dotted lines stand for $N=N_{A}=N_{B}=200$ and for $N=2000$
with couplings corresponding to $B=9.116\,\mathrm{G}$; while the
dash-dotted and dashed lines assume no local cross couplings, i.e.,
$g_{12}=0$, $g_{11}=g_{22}$, for $N=200$ and for $N=2000$.\label{fig:Entanglement} }}
\end{figure}

After using the beam splitter, entanglement can be detected as $E_{\mathrm{product}}<1$,
as shown in Fig.~\ref{fig:Entanglement} by the solid curve, which
assumes the couplings between spins found at the rubidium Feshbach
resonance. Note that, consistent with the results found in previous
studies of entanglement in the ground state, the dashed curve of Fig.~\ref{fig:Entanglement}
shows that no cross couplings, i.e., $g_{12}=0$, gives much better
results still. This would require spatially separated condensates
for each spin orientation, in order to eliminate cross couplings,
as recently demonstrated by using magnetic gradient techniques~\cite{Philipp2010}.
\textcolor{black}{Fig.~\ref{fig:Entanglement} reveals improvement
in the entanglement, as one increases the number of atoms in the condensate.}

\subsection{EPR \textcolor{black}{paradox }entanglement\label{sec:EPRparadox}}

The entanglement predicted for the dynamical scheme is strong enough
that it reveals an EPR paradox (steering) non-locality. This level
of entanglement is reached when $E_{\mathrm{product}}<0.5$ (as shown
in Section~\ref{sec:Measurement-strategies}), which occurs in the
presence of local couplings for the larger atom numbers $N=2000$,
as shown in Fig.~\ref{fig:Entanglement}. The EPR paradox entanglement
can be obtained for lower atom numbers $N\sim200$ when local couplings
are non-existent.

Next, we will examine the predictions for the more sensitive EPR criterion
Eq.~(\ref{eq:eprcritpordxp}), which involves measurements of the
conditional variances. The EPR argument is based on an accuracy of
inference, that an observer at $D$ can predict the result $\hat{J}_{C}^{\theta}$
for an observer at $C$, to a certain measurable level of uncertainty.
A simple way to determine an upper limit to this uncertainty is to
use a linear estimate $g\hat{J}_{D}^{\theta}$, based on the result
$\hat{J}_{D}^{\theta}$ for measurement at $D$. Then we arrive at
the EPR paradox criterion of Eq.~(\ref{eq:product form epr-1}):
\begin{eqnarray}
E_{\mathrm{EPR-product}} & = & \frac{\sqrt{\Delta^{2}(\hat{J}_{C}^{\theta}-g\hat{J}_{D}^{\theta})\cdot\Delta^{2}(\hat{J}_{C}^{\theta+\pi/2}+g^{\prime}\hat{J}_{D}^{\theta+\pi/2})}}{|\langle\hat{J}_{C}^{Y}\rangle|/2}\nonumber \\
 & < & 1,\label{eq:product form epr}
\end{eqnarray}
which reduces to $ $ $E_{\mathrm{product}}<0.5$, for the choice
$g=1$. In fact, this choice of $g$ is optimal where the inference
squeezing is very strong. 

\begin{figure}[t]
\begin{centering}
\includegraphics[width=1\columnwidth]{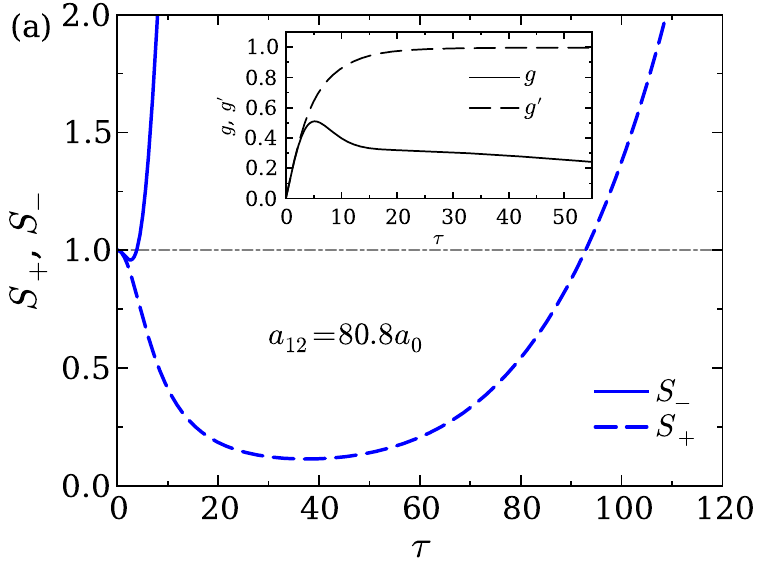}
\includegraphics[width=1\columnwidth]{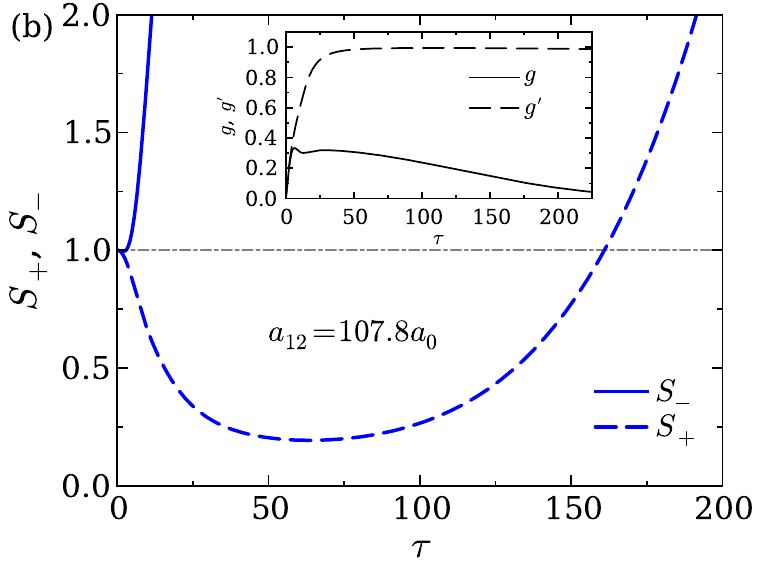}
\par\end{centering}

\centering{}\caption{EPR paradox entanglement, shown by the simultaneous \textcolor{black}{inference
squeezing of $S_{-}=\left[\Delta^{2}(\hat{J}_{C}^{\theta}-g\hat{J}_{D}^{\theta})/n_{0}\right]$
(solid lines) and $S_{+}=\left[\Delta^{2}(\hat{J}_{C}^{\theta+\pi/2}+g^{\prime}\hat{J}_{D}^{\theta+\pi/2})/n_{0}\right]$
(dashed lines)}\textcolor{red}{{} }\textcolor{black}{with optimal $g$
and $g^{\prime}$, where the shot noise level is $n_{0}=(|\langle\hat{J}_{C}^{Y}\rangle|)/2$}.
Plots show different local cross couplings: (a) $B=9.116\,\mathrm{G}$,
and (b) $B=9.086\,\mathrm{G}$\textcolor{black}{. Insets show the
optimal value of factors $g$ (solid lines) and $g^{\prime}$ (dashed
lines) with time. Here $N_{A}=N_{B}=200$, and oth}er parameters are
as for Fig.~\ref{fig:squeezing-after-BEC.}.\label{fig:inference-squeezing n=00003D200} }
\end{figure}

We now determine how to optimize the choice of $g$, where the entanglement
is weaker. The best choices for $g$ and $g'$ are adjusted to minimize
$\Delta^{2}(\hat{J}_{C}^{\theta}-g\hat{J}_{D}^{\theta})$ and $\Delta^{2}(\hat{J}_{C}^{\theta+\pi/2}+g'\hat{J}_{D}^{\theta+\pi/2})$.
Following~\cite{eprcrit},
\begin{eqnarray}
\frac{\partial}{\partial g}\Delta^{2}(\hat{J}_{C}^{\theta}-g\hat{J}_{D}^{\theta}) & = & 2g\Delta^{2}\hat{J}_{D}^{\theta}-\langle\hat{J}_{C}^{\theta},\hat{J}_{D}^{\theta}\rangle-\langle\hat{J}_{D}^{\theta},\hat{J}_{C}^{\theta}\rangle\nonumber \\
 & = & 0
\end{eqnarray}
implies the optimal $g$ is given by
\begin{equation}
2g=(\langle\hat{J}_{C}^{\theta},\hat{J}_{D}^{\theta}\rangle+\langle\hat{J}_{D}^{\theta},\hat{J}_{C}^{\theta}\rangle)/\Delta^{2}\hat{J}_{D}^{\theta}.\label{eq:optimal g}
\end{equation}
We note $\langle\hat{J}_{C}^{\theta},\hat{J}_{D}^{\theta}\rangle=\langle\hat{J}_{D}^{\theta},\hat{J}_{C}^{\theta}\rangle$.
There is similarly an optimum for the value of $g^{\prime}$, with
phase $\theta^{\prime}=\theta+\pi/2$. Figure~\ref{fig:inference-squeezing n=00003D200}
shows the optimal value of factor $g$ and $g^{\prime}$ with different
cross couplings, versus time, for atoms $N_{A}=200$.

Using the optimal values of the constants, one can evaluate the predictions
for the variances, $\Delta^{2}(\hat{J}_{C}^{\theta}-g\hat{J}_{D}^{\theta})$.
We can express the required moments in terms of the moments of modes
$a$, $b$, e.g.,
\begin{eqnarray}
\hat{J}_{C}^{Z}-g\hat{J}_{D}^{Z} & = & g_{-}(\hat{J}_{A}^{Z}+\hat{J}_{B}^{Z})\\
 &  & +ig_{+}[\frac{\hat{a}_{2}^{\dagger}\hat{b}_{2}-\hat{b}_{2}^{\dagger}\hat{a}_{2}-\hat{a}_{1}^{\dagger}\hat{b}_{1}+\hat{b}_{1}^{\dagger}\hat{a}_{1}}{2}],\nonumber 
\end{eqnarray}
where we define $g_{\pm}=\left(1\pm g\right)/2$, and use $\theta_{2}-\phi_{2}=\theta_{1}-\phi_{1}=0$
due to the symmetry of $a$ and $b$. The minimum variance is obtained
by substitution of optimal $g$, Eq.~(\ref{eq:optimal g}). One finds
\begin{eqnarray}
\Delta^{2}(\hat{J}_{C}^{Z}-g\hat{J}_{D}^{Z}) & = & g_{-}^{2}\Delta^{2}\hat{J}_{+}^{Z}+g_{+}^{2}\Delta^{2}\hat{J}_{-}^{Z},
\end{eqnarray}
where
\begin{eqnarray}
\Delta^{2}\hat{J}_{-}^{Z} & = & \Delta^{2}(\hat{J}_{C}^{Z}-\hat{J}_{D}^{Z})\nonumber \\
 & = & \Delta^{2}(i\frac{\hat{a}_{2}^{\dagger}\hat{b}_{2}-\hat{b}_{2}^{\dagger}\hat{a}_{2}-\hat{a}_{1}^{\dagger}\hat{b}_{1}+\hat{b}_{1}^{\dagger}\hat{a}_{1}}{2}),\nonumber \\
\Delta^{2}\hat{J}_{+}^{Z} & = & \Delta^{2}(\hat{J}_{C}^{Z}+\hat{J}_{D}^{Z})=\Delta^{2}(\hat{J}_{A}^{Z}+\hat{J}_{B}^{Z}).
\end{eqnarray}
Similarly
\begin{eqnarray}
\hat{J}_{C}^{X}-g\hat{J}_{D}^{X} & = & g_{-}(\hat{J}_{A}^{X}+\hat{J}_{B}^{X})\\
 &  & -g_{+}[\frac{\hat{a}_{2}^{\dagger}\hat{b}_{1}-\hat{b}_{2}^{\dagger}\hat{a}_{1}+\hat{a}_{1}^{\dagger}\hat{b}_{2}-\hat{b}_{1}^{\dagger}\hat{a}_{2}}{2}]\nonumber 
\end{eqnarray}
with $\theta_{2}-\phi_{1}=-(\theta_{1}-\phi_{2})=\pi/2$, for which
the minimum variance is
\begin{eqnarray}
\Delta^{2}(\hat{J}_{C}^{X}-g\hat{J}_{D}^{X}) & = & g_{-}^{2}\Delta^{2}\hat{J}_{+}^{X}+g_{+}^{2}\Delta^{2}\hat{J}_{-}^{X},
\end{eqnarray}
where
\begin{eqnarray}
\Delta^{2}\hat{J}_{-}^{X} & = & \Delta^{2}(\hat{J}_{C}^{X}-\hat{J}_{D}^{X})\nonumber \\
 & = & \Delta^{2}(-\frac{\hat{a}_{2}^{\dagger}\hat{b}_{1}-\hat{b}_{2}^{\dagger}\hat{a}_{1}+\hat{a}_{1}^{\dagger}\hat{b}_{2}-\hat{b}_{1}^{\dagger}\hat{a}_{2}}{2}),\nonumber \\
\Delta^{2}\hat{J}_{+}^{X} & = & \Delta^{2}(\hat{J}_{C}^{X}+\hat{J}_{D}^{X})=\Delta^{2}(\hat{J}_{A}^{X}+\hat{J}_{B}^{X}).
\end{eqnarray}
Also, 
\begin{eqnarray}
\langle\hat{J}_{C}^{Z}-g\hat{J}_{D}^{Z},\hat{J}_{C}^{X}-g\hat{J}_{D}^{X}\rangle & = & g_{-}^{2}\langle\hat{J}_{+}^{Z},\hat{J}_{+}^{X}\rangle\nonumber \\
 &  & +g_{+}^{2}\langle\hat{J}_{-}^{Z},\hat{J}_{-}^{X}\rangle.
\end{eqnarray}
The minimum interference squeezing $\Delta^{2}(\hat{J}_{C}^{\theta}-g\hat{J}_{D}^{\theta})$
and $\Delta^{2}(\hat{J}_{C}^{\theta+\pi/2}+g^{\prime}\hat{J}_{D}^{\theta+\pi/2})$
with optimal choices of $g$ and $g^{\prime}$ are shown in Fig.~\ref{fig:inference-squeezing n=00003D200}.
Unlike the original formulations of the EPR paradox, in this case,
the two inference variances are asymmetric. The second inference variance
exceeds the quantum limit for large enough $\tau$. We note that for
strong correlation as shown by $S_{+}$ over a large range of $\tau$
the optimal choice for $g$ becomes $1$. On the other hand, for the
poor correlation shown by $S_{-}$ as $\tau$ becomes larger, the
optimal choice becomes $g^{\prime}=0$, so that the variance $S_{+}$
is limited to the variance of Bob's spin (Fig.~\ref{fig:inference-squeezing n=00003D200}).

\begin{figure}
\begin{centering}
\includegraphics[width=1\columnwidth]{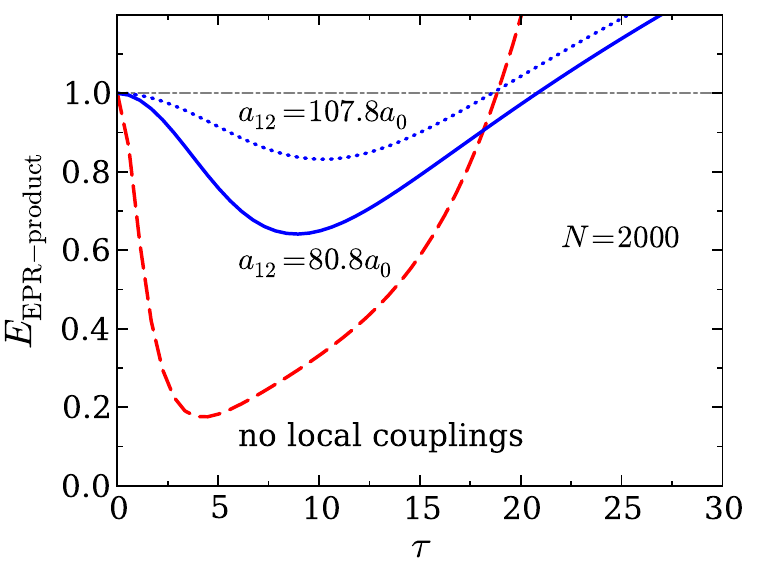}
\par\end{centering}

\caption{EPR paradox is predicted ($E_{\mathrm{EPR-product}}<1$), for a v\textcolor{black}{ariety
of local cross couplings. Here $N=N_{A}=N_{B}=2000$; }$B=9.116\,\mathrm{G}$\textcolor{black}{{}
(solid line), }$B=9.086\,\mathrm{G}$\textcolor{black}{{} (dotted line),
and with no cross-couplings, i.e., $g_{12}=0,\ g_{22}=g_{11}$ (dashed
line). We use optimal value of factors $g$ and $g'$ for each case.
Other parameters are as for the last figures.\label{fig:EPR-paradox} }}
\end{figure}

\begin{figure}[h]
\begin{centering}
\includegraphics[width=1\columnwidth]{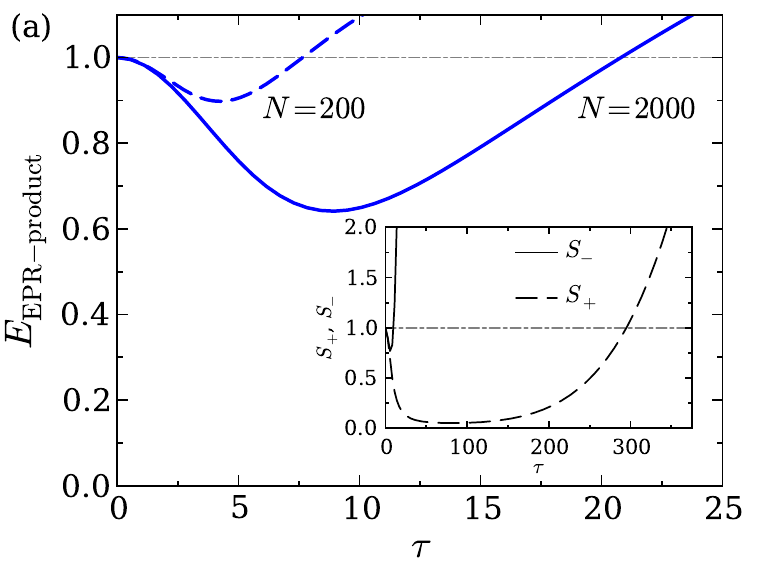} \includegraphics[width=1\columnwidth]{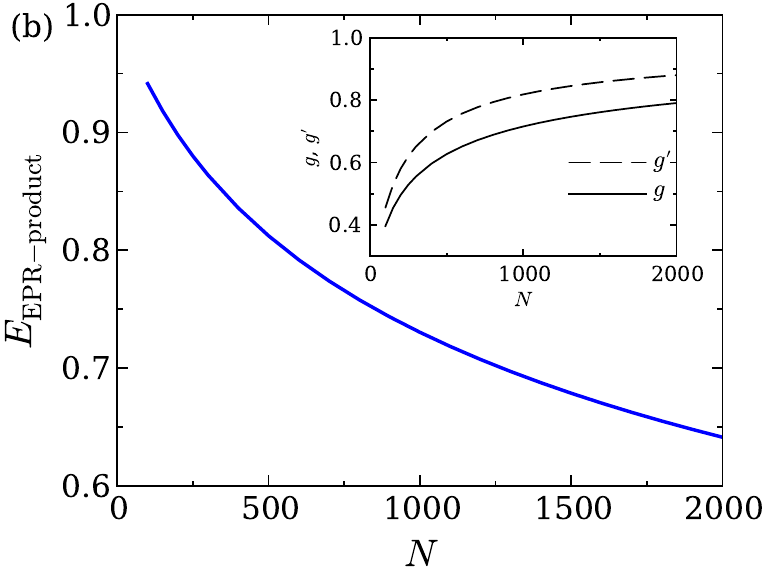}
\par\end{centering}

\caption{Effect of atom number: (a) $N=N_{A}=N_{B}=200$ (dashed lines) and
$N=2000$ (solid lines) on the EPR paradox entanglement with optimal
gain factors $g$ and $g^{\prime}$, for fixed couplings $B=9.116\,\mathrm{G}$.
Inset shows the individual squeezing inferences $S_{-}=\left[\Delta^{2}(\hat{J}_{C}^{\theta}-g\hat{J}_{D}^{\theta})/n_{0}\right]$
(solid lines), $S_{+}=\left[\Delta^{2}(\hat{J}_{C}^{\theta+\pi/2}+g^{\prime}\hat{J}_{D}^{\theta+\pi/2})/n_{0}\right]$
(dashed lines) with optimal $g$ and $g^{\prime}$, for $N=2000$.\textcolor{red}{{}
}(b) The optimal $E_{\mathrm{EPR-product}}$ versus different number
of atoms. Inset shows the corresponding $g$, $g^{\prime}$ versus
$N$.\label{fig:Effect-of-atom-number}}
\end{figure}

Figure~\ref{fig:EPR-paradox} shows regimes for which the EPR criteria
are satisfied, i.e., $E_{\mathrm{EPR-product}}<1$, for different
cross couplings. As the number of atoms increases, the spin squeezing
increases and so too does the degree of EPR paradox, as shown in Fig.~\ref{fig:Effect-of-atom-number}.
The result is consistent with previous studies of the CV EPR paradox
non-locality~\cite{ou,RDeprlarge}. An EPR paradox for the quadrature
phase amplitudes of optical modes has been confirmed in a number of
experiments~\cite{rmp}. Whether this effect is realizable for atoms
however is a different question. Considerations include the size of
$\tau$ compared to the decoherence time of the BEC condensate and
whether interactions like nonlinear losses ignored in the model~(\ref{eq:ham1})
will come into play to reduce the EPR entanglement. To address these
questions, we employ the stochastic truncated Wigner function technique
valid for large $N$~\cite{wigner}, in the next section.

\section{Strategy II: Simultaneous evolution with tunneling present and including
losses\label{sec:Truncated-Wigner-application}}

The degree of EPR entanglement is limited by the number of atoms ($N$)
in the ensemble. Figutre~\ref{fig:Effect-of-atom-number} shows a
base value of $\sim0.83$ for $N=200$, but the entanglement improves
to $\sim0.65$ for $N=2000$. However, the unitary evolution approach
of the previous section becomes limited once dissipation effects are
important. These effects are known to occur in ultra-cold atomic systems
especially at high densities, due to spin-changing atomic collisions
which cause density-dependent two-body and three-body losses. With
this in mind we investigate the effectiveness of using a large number
approximation, namely the truncated Wigner function technique~\cite{wigner}.
In this approach, higher order terms in $1/N^{3/2}$ are ignored,
to allow a stochastic calculation based on a positive Wigner function.
This method readily scales to large numbers of atoms and modes~\cite{Egorov},
and can include nonlinear losses. The detailed description of the
method can be found in Appendix~\ref{sec:Truncated-Wigner-Method}.

\begin{figure}[t]
\begin{centering}
\includegraphics[width=1\columnwidth]{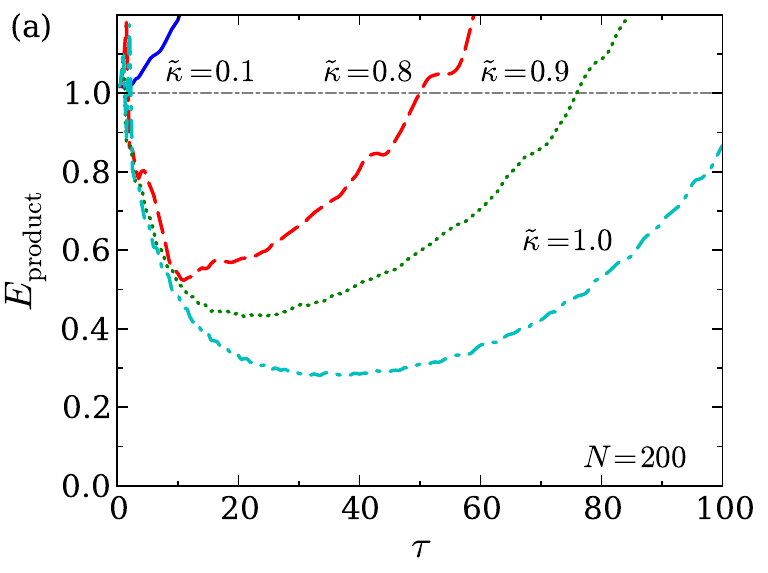}
\includegraphics[width=1\columnwidth]{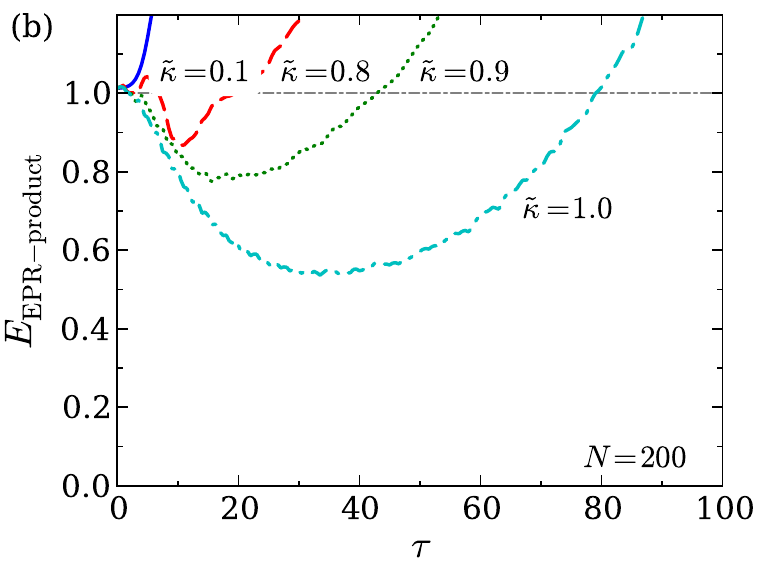} 
\par\end{centering}

\caption{(Color online) Entanglement $E_{\mathrm{product}}$ (a) and $E_{\mathrm{EPR-product}}$
(b) after evolution of the full Hamiltonian~(\ref{eq:Hamiltonian})
including tunneling for $N=200$ and $B=9.116\,\mathrm{G}$, without
the application of the beam splitter. The results are shown for values
$\tilde{\kappa}=0.1$ (blue solid lines), $\tilde{\kappa}=0.8$ (red
dashed lines), $\tilde{\kappa}=0.9$ (green dotted lines) and $\tilde{\kappa}=1$
(cyan dash-dotted lines).\label{fig:w:kappa_effect}}
\end{figure}

The predictions of the truncated Wigner method are indistinguishable
from the exact method given in Appendix~\ref{sec:Dynamical-solution}
in the case of zero losses and no tunneling, which confirms its validity
for $N\sim200\text{\textendash}2000$. The advantage of the Wigner
method is that it allows a ready solution of the dynamics of the full
Hamiltonian~(\ref{hamgs-1-1},~\ref{eq:Hamiltonian}) where both
tunneling and nonlinear terms are present, even for large atom numbers.
The method also allows the inclusion of losses, which are known to
destroy squeezing and entanglement, and will come into play in realistic
experimental arrangements. For example, tunneling cannot be completely
suppressed, and also potentially significant is the role of nonlinear
loss, that will come into play with large atom numbers. 

\begin{figure}
\begin{centering}
\includegraphics[width=1\columnwidth]{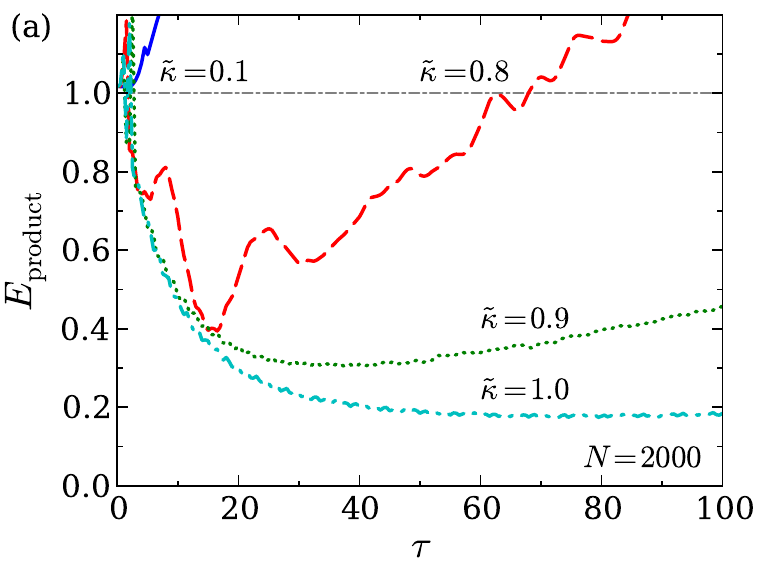}
\includegraphics[width=1\columnwidth]{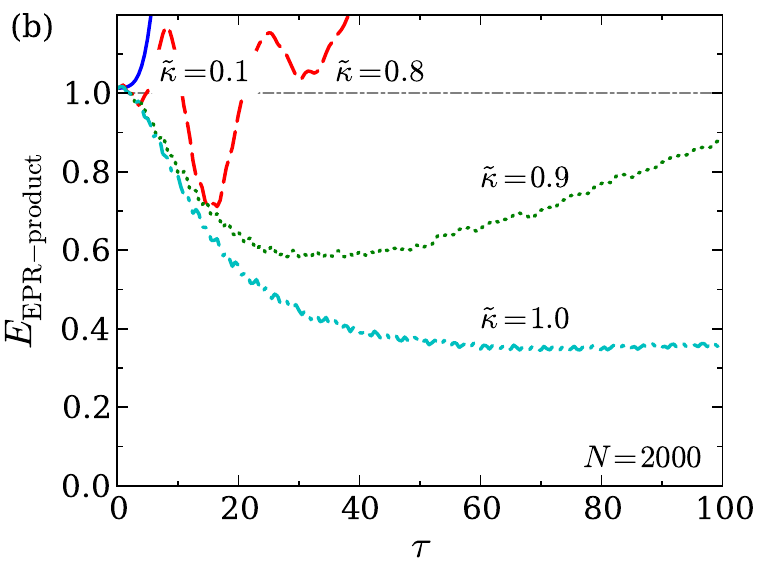} 
\par\end{centering}

\caption{(Color online) Entanglement $E_{\mathrm{product}}$ (a) and $E_{\mathrm{EPR-product}}$
(b) after evolution of the full Hamiltonian~(\ref{eq:Hamiltonian})
including tunneling for $N=2000$ and $B=9.116\,\mathrm{G}$. Meaning
of the lines is the same as in Fig.~\ref{fig:w:kappa_effect}.\label{fig:w:kappa_effect_2k}}
\end{figure}

Once tunneling is present in the nonlinear Hamiltonian, entanglement
can be created between the two modes, \emph{without} the second beam
splitter step described in Section~\ref{sec:Two-step-dynamical-entanglement}.
Figure~\ref{fig:w:kappa_effect} shows the effect of tunneling on
the entanglement. Strong tunneling ($\tilde{\kappa}\sim1$, i.e.,
with the strength of the same order as the nonlinear interaction)
produces significant entanglement even without the final application
of the beam splitter. However, tunneling this strong is hard to achieve
in a simple two-well BEC experiment, where values of $\tilde{\kappa}\sim10^{-3}\ldots10^{-2}$
are more common. The achievement of such large couplings is likely
to require a more sophisticated experimental design.

If one increases the number of atoms $N$, with the absence of losses
the dimensionless drift part of equations~(\ref{eq:w:sdes}) will
stay the same, but the results will differ, as shown in Fig.~\ref{fig:w:kappa_effect_2k}.
The entanglement improves with higher $N$, as in the case of the
two-step strategy and as found previously for ground state calculations.
This result is consistent for continuous variable EPR entanglement,
which has been predicted in optics for high intensity Gaussian states.

\begin{figure}
\begin{centering}
\includegraphics[width=1\columnwidth]{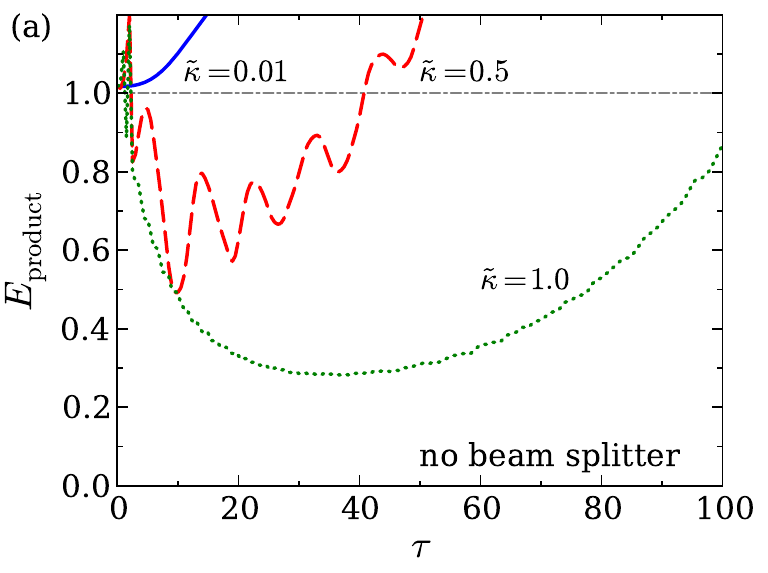}
\includegraphics[width=1\columnwidth]{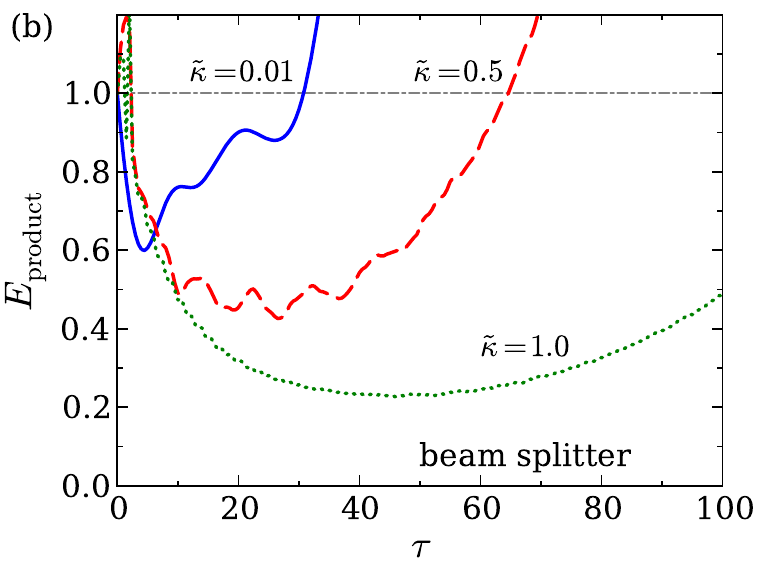} 
\par\end{centering}

\caption{(Color online) Entanglement $E_{\mathrm{product}}$ after evolution
of the full Hamiltonian (\ref{eq:Hamiltonian}) including tunneling
for $N=200$ and $B=9.116\,\mathrm{G}$, (a) without and (b) with
the application of the beam splitter. The results are shown for values
$\tilde{\kappa}=0.01$ (blue solid lines), $\tilde{\kappa}=0.5$ (red
dashed lines), and $\tilde{\kappa}=1$ (green dotted lines).\label{fig:w:beam_splitter_effect}}
\end{figure}

We also analyze the predictions for entanglement with the insertion
of step $2$, the ``beam splitter'' interaction described in Section
\ref{sec:Two-step-dynamical-entanglement}.B, the objective being
to understand whether a small amount of tunneling and nonlinear loss
present in the first interaction stage will detract from the amount
of entanglement that is predicted by the two-step strategy. We focus
on the case where local couplings are present, although this gives
a worse case prediction for entanglement. The losses and tunneling
tend to detract from the amount of entanglement, so that in the case
of $2000$ atoms, with local couplings, an entanglement of $\sim0.4$
(enough to give an EPR paradox) as shown in Fig.~\ref{fig:Entanglement}
is reduced to $\sim0.6$ (Fig.~\ref{fig:w:beam_splitter_effect}).
We emphasize here that losses are highly adjustable through changes
in density, so these issues are more related to appropriate experimental
design than to fundamental limits. The application of the beam splitter
improves entanglement significantly when tunneling is weak, but with
$\tilde{\kappa}$ values close to 1 tunneling becomes the prevalent
source of entanglement, as illustrated in Fig.~\ref{fig:w:beam_splitter_effect}.

\begin{figure}
\begin{centering}
\includegraphics[width=1\columnwidth]{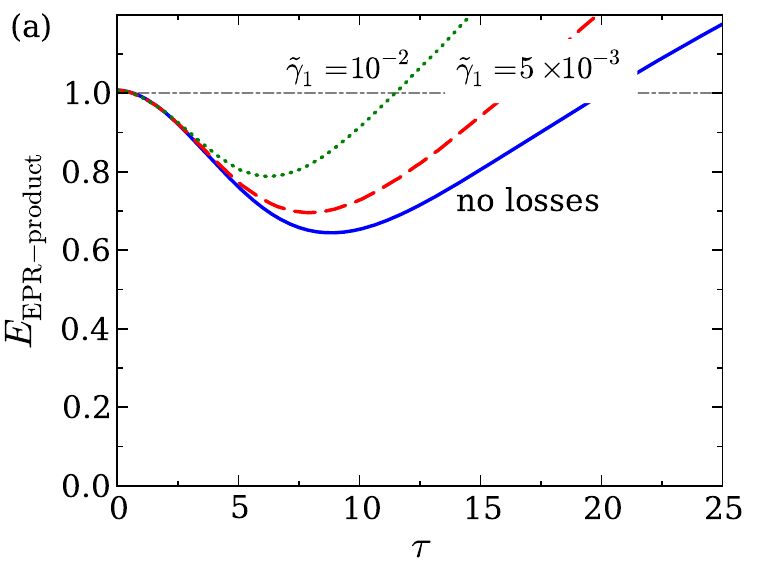}
\includegraphics[width=1\columnwidth]{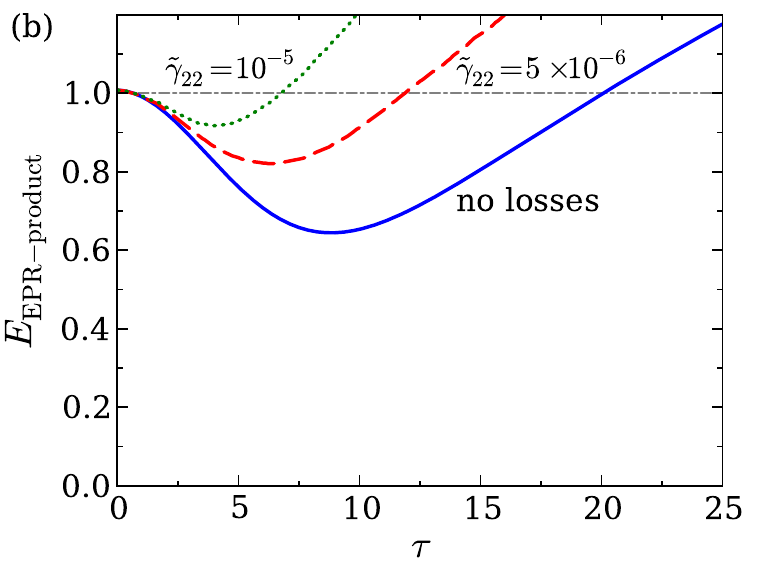}
\par\end{centering}

\caption{(Color online) EPR paradox entanglement as measured by $E_{\mathrm{EPR-product}}$
without tunneling and with the application of the beam splitter, for
$B=9.116\,\mathrm{G}$ and $N_{A}=N_{B}=2000$. (a) Only linear losses
are enabled; no losses (solid blue lines), $\tilde{\gamma}_{1}=5\times10^{-3}$
(dashed red lines), $\tilde{\gamma}_{1}=10^{-2}$ (dotted green lines).
(b) Only two-body intra-species losses are enabled; no losses (solid
blue lines), $\tilde{\gamma}_{22}=5\times10^{-6}$ (dashed red lines),
$\tilde{\gamma}_{22}=10^{-5}$ (dotted green lines).\label{fig:w:losses_effect_normal}}
\end{figure}

\begin{figure}
\begin{centering}
\includegraphics[width=1\columnwidth]{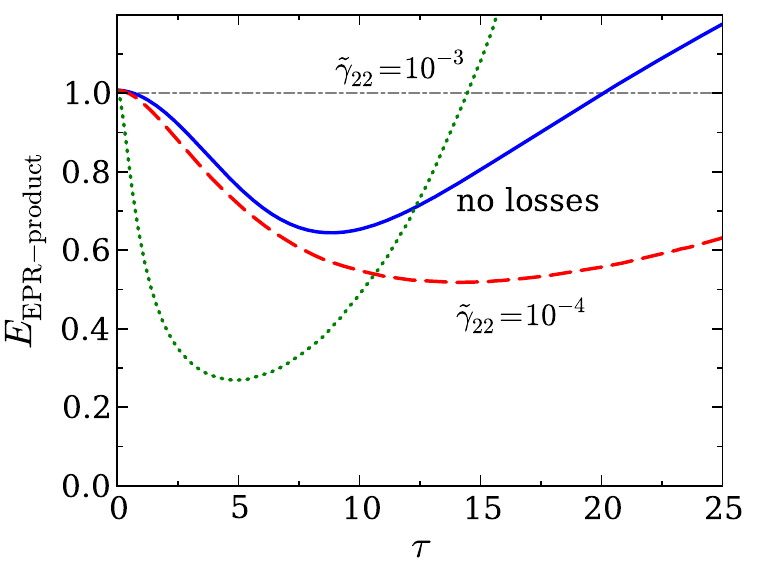}
\par\end{centering}

\caption{(Color online) EPR paradox entanglement as measured by $E_{\mathrm{EPR-product}}$
without tunneling and with the application of the beam splitter, for
$B=9.116\,\mathrm{G}$ and $N_{A}=N_{B}=2000$. Only two-body inter-species
losses are enabled; no losses (solid blue lines), $\tilde{\gamma}_{12}=10^{-4}$
(dashed red lines), $\tilde{\gamma}_{12}=10^{-3}$ (dotted green lines).\label{fig:w:losses_effect_interspecies}}
\end{figure}

The presence of linear losses, or intra-species losses decreases the
maximum entanglement, as shown in Fig.~\ref{fig:w:losses_effect_normal}.
But when the nonlinear inter-species losses are enabled, our simulations
show that they unexpectedly \emph{increase} the entanglement (Fig.~\ref{fig:w:losses_effect_interspecies}). 

We have found that this only occurs with a nonlinear loss $\tilde{\gamma}_{12}$
that specifically couples the two species together. Other forms of
loss, including linear loss, will simply reduce the entanglement \textemdash{}
as one expects from decoherence effects. Generating entanglement by
a manipulation of reservoirs has been observed for atomic ensembles~\cite{engres}.
The process in our case appears to be related to the fact that nonlinear
absorption just by itself is known to create a nonclassical state~\cite{Chaturvedi1977},
which can then become entangled through linear couplings alone.

\section{Summary\label{sec:Summary}}

In summary, we have analyzed two strategies for dynamical preparation
of squeezed and entangled atomic states, through the use of nonlinear
interactions that rely on the naturally occurring S-wave interactions
between trapped atoms. The possible advantage of the methods proposed
here is that they do not require the use of separate local oscillators,
as often employed in optics for measuring entanglement. These could
be potentially technically difficult to use in some cold-atom systems,
due to dephasing that is caused by interatomic interactions in the
local-oscillator itself, when combined with number fluctuatio\textcolor{black}{ns.
However, atomic homodyning has recently been realised \cite{neweprbec2grossent},
and may present a better strategy in some cases, particularly where
it is important to avoid local cross-couplings. Work done in parallel
with ours studies the dynamical prepration of entanglement for this
case, using a single mode at each well \cite{murray dynamical}.}

\textcolor{black}{We find that robust spatial entanglement and EPR
inference is possible under the correct conditions, even including
Poissonian number fluctuations. These effects are also not greatly
perturbed by realistic loss values. In fact, with the presence of
certain types of inter-species loss, we calculate that enhanced entanglement
is possible. Finally, we note that the optimum regime is for rather
large coupling or tunneling values between the spatial modes, which
appears to require a different experimental design to a simple two-well
system.}

\textcolor{black}{The EPR paradox entanglement studied in this paper
confirms an inconsistency of local realism with the completeness of
quanutm mechanics \cite{epr}, and is evidence for the form of nonlocality
called ``steering'' \cite{hw-steering-1}. We make the point however
that EPR entanglement is not itself sufficient to imply a direct failure
of local realism, as would be demonstrated by a violation of a Bell
inequality \cite{Bell,ou}. The method employed to arrive at the predictions
of EPR entanglement illustrates this point, since the truncated Wigner
function is positive, and therefore provides a local hidden variable
theory to describe the statistics of experiments where the measurement
is effectively that of a quadrature phase amplitude, even though significant
EPR entanglement can be obtained \cite{ou}. The distinction between
the EPR steering and Bell forms of nonlocality has been emphasised
further in recent papers \cite{EPRsteering-1}. }
\begin{acknowledgments}
We wish to thank the Australian Research Council for funding via ACQAO
COE, Discovery and DECRA grants, as well as useful discussions with
M. Oberthaler, P. Treutlein and A. Sidorov.
\end{acknowledgments}
\appendix

\section{Exact dynamical solutions\label{sec:Dynamical-solution}}

\subsubsection{One-mode model: }

We start by considering a simple nonlinear, single-mode interaction
that occurs locally at each well, and is modeled by Hamiltonian
\begin{equation}
\hat{H}/\hbar=\frac{g}{2}:\hat{N}^{2}:.
\end{equation}

In order to understand the dynamics induced by the above Hamiltonian,
we calculate results in the Heisenberg picture, where one obtains:
\begin{eqnarray}
\frac{d\hat{a}_{i}}{dt} & = & \frac{i}{\hbar}\left[\hat{H},\hat{a}_{i}\right]=-ig\hat{a}^{\dagger}\hat{a}\hat{a}=-ig\hat{N}\hat{a}.
\end{eqnarray}
Since the number of particles is conserved, this has the solution:
\begin{eqnarray}
\hat{a}\left(t\right) & = & \exp\left[-ig\hat{N}t\right]\hat{a}\left(0\right),
\end{eqnarray}
which gives
\begin{eqnarray}
\left\langle \hat{a}\left(t\right)\right\rangle  & = & \left\langle \Psi(0)\right|\hat{a}\left(t\right)\left|\Psi(0)\right\rangle \nonumber \\
 & = & \sum_{n,m=0}^{\infty}C_{n}^{*}C_{m}\left\langle n\right|\exp\left[-ig\hat{N}t\right]\hat{a}\left(0\right)\left|m\right\rangle \nonumber \\
 & = & \sum_{n=0}^{\infty}C_{n}^{*}C_{n+1}\sqrt{n+1}\exp\left[-ignt\right]\nonumber \\
 & = & \alpha e^{-|\alpha|^{2}}\sum_{n=0}^{\infty}e^{-ignt}\frac{|\alpha|^{2n}}{n!}\nonumber \\
 & = & \alpha\exp\left[|\alpha|^{2}\left(e^{-igt}-1\right)\right].
\end{eqnarray}

This predicts a well-known behavior with three characteristic time
scales:
\begin{itemize}
\item On very short time scales, there is simply an oscillation with a renormalized
frequency of $\omega'=g\overline{n}$, where $\overline{n}=|\alpha|^{2}$
. A similar result is obtained classically. 
\item On intermediate time-scales, there is a quadratic damping, with a
characteristic damping rate of $g\sqrt{\overline{n}}$, which is proportional
to the standard deviation in the initial particle number. 
\item Finally, on very long time-scales there is a succession of periodic
revivals where the initial state and all its properties are regained
exactly, apart from a possible phase-shift. This occurs whenever $t=2\pi/g$
.
\end{itemize}

\subsubsection{Two-mode model:}

Next, consider a simple two-mode model, with two internal (spin) modes,
where the Hamiltonian for the coupled system is:
\begin{eqnarray}
\hat{H} & = & \frac{\hbar}{2}\sum_{ij}g_{ij}\hat{a}_{i}^{\dagger}\hat{a}_{j}^{\dagger}\hat{a}_{j}\hat{a}_{i}=\frac{\hbar}{2}:\sum_{ij}g_{ij}\hat{N}_{i}\hat{N}_{j}:.\label{eq:nonlinear H-1}
\end{eqnarray}

We can solve this using either Schroedinger or Heisenberg equations
of motion. In the Heisenberg case, one obtains:
\begin{eqnarray}
\frac{d\hat{a}_{i}}{dt} & = & \frac{i}{\hbar}\left[\hat{H},\hat{a}_{i}\right]=-i\sum_{j}g_{ij}\hat{a}_{j}^{\dagger}\hat{a}_{j}\hat{a}_{i}=-i\sum_{j}g_{ij}\hat{N}_{j}\hat{a}_{i}.\nonumber \\
\end{eqnarray}
Since the number of particles is conserved in each mode, this has
the solution:
\begin{eqnarray}
\hat{a}_{i}\left(t\right) & = & \exp\left[-i\sum_{j}g_{ij}\hat{N}_{j}t\right]\hat{a}_{i}\left(0\right).
\end{eqnarray}

We suppose the initial quantum state factorizes into a vacuum state
in one mode and coherent state in the second, and that the condensate
mode is give by:
\begin{eqnarray}
\left|\Psi(0)\right\rangle  & = & \left|0\right\rangle _{a_{1}}\left|\alpha\right\rangle _{a_{2}}=\left|0\right\rangle _{a_{1}}\sum_{n=0}^{\infty}C_{n}\left|n\right\rangle _{a_{2}},
\end{eqnarray}
which gives 
\begin{eqnarray}
\hat{a}_{1}^{\dagger}\hat{a}_{1}\left|\Psi(0)\right\rangle  & = & 0,\nonumber \\
\hat{a}_{2}^{\dagger}\hat{a}_{2}\left|\Psi(0)\right\rangle  & = & \left|\alpha\right|^{2}\left|\Psi(0)\right\rangle .
\end{eqnarray}
In this coherent state the number fluctuation variance is $N$. Suppose
we apply a linear beamsplitter, then, after the beam-splitter:
\begin{eqnarray}
\bar{a}_{1} & = & \frac{1}{\sqrt{2}}(\hat{a}_{1}+\hat{a}_{2}),\nonumber \\
\bar{a}_{2} & = & \frac{1}{\sqrt{2}}(\hat{a}_{2}-\hat{a}_{1}),
\end{eqnarray}
the state becomes
\begin{equation}
|\bar{\Psi}(0)\rangle=|\frac{\alpha}{\sqrt{2}}\rangle_{a_{1}}|\frac{\alpha}{\sqrt{2}}\rangle_{a_{2}}.
\end{equation}

After the application of the nonlinear Hamiltonian (\ref{eq:nonlinear H-1}),
squeezing is generated locally, and
\begin{eqnarray}
\left\langle \bar{a}_{i}(t)\right\rangle  & = & \langle\bar{\Psi}(0)|\bar{a}_{i}(t)|\bar{\Psi}(0)\rangle\nonumber \\
 & = & \langle\frac{\alpha}{\sqrt{2}}|\langle\frac{\alpha}{\sqrt{2}}|\exp\left[-i\sum_{j}g_{ij}\hat{N}_{j}t\right]\bar{a}_{i}|\frac{\alpha}{\sqrt{2}}\rangle|\frac{\alpha}{\sqrt{2}}\rangle\nonumber \\
 & = & \sum_{n_{1},n_{2},m_{1},m_{2}=0}^{\infty}C_{n_{1}}^{*}C_{n_{2}}^{*}C_{m_{1}}C_{m_{2}}\nonumber \\
 &  & \times\langle n_{1}|\langle n_{2}|\exp\left[-i\sum_{j}g_{ij}\hat{N}_{j}t\right]\bar{a}_{i}|m_{1}\rangle|m_{2}\rangle\nonumber \\
 & = & \frac{\alpha}{\sqrt{2}}\exp\left[\frac{|\alpha|^{2}}{2}\left(e^{-ig_{i1}t}-1\right)\right]\exp\left[\frac{|\alpha|^{2}}{2}\left(e^{-ig_{i2}t}-1\right)\right].\nonumber \\
\end{eqnarray}

We wish to calculate the phase variance in $\hat{\theta}\equiv\hat{\theta}_{1}-\hat{\theta}_{2}$,
as this gives rise to the decay in an interference pattern:
\begin{equation}
\langle(\Delta\hat{\theta})^{2}\rangle\equiv\langle\hat{\theta}^{2}\rangle-\langle\hat{\theta}\rangle^{2}.
\end{equation}

\section{Truncated Wigner Method\label{sec:Truncated-Wigner-Method}}

Including losses, the master equation in four-mode approximation is
written as
\begin{equation}
\frac{d\hat{\rho}}{dt}=-\frac{i}{\hbar}[\hat{H},\hat{\rho}]+\sum_{\bm{l}}\gamma_{\bm{l}}\mathcal{L}_{\bm{l}}[\hat{\rho}],\label{eq:w:master-eqn}
\end{equation}
where the Hamiltonian is defined as~(\ref{eq:Hamiltonian}), and
loss term has the form 
\begin{equation}
\mathcal{L}_{\bm{l}}[\hat{\rho}]=2\hat{O}_{\bm{l}}\hat{\rho}\hat{O}_{\bm{l}}^{\dagger}-\hat{O}_{\bm{l}}^{\dagger}\hat{O}_{\bm{l}}\hat{\rho}-\hat{\rho}\hat{O}_{\bm{l}}^{\dagger}\hat{O}_{\bm{l}}.\label{eq:w:loss-term}
\end{equation}
We consider three different sources of losses: $\hat{O}_{22A}=\hat{a}_{2}^{2}$
and $\hat{O}_{22B}=\hat{b}_{2}^{2}$ (two-body intra-species loss),
$\hat{O}_{12A}=\hat{a}_{1}\hat{a}_{2}$ and $\hat{O}_{12B}=\hat{b}_{1}\hat{b}_{2}$
(two-body inter-species loss), $\hat{O}_{1A}=\hat{a}_{2}$ and $\hat{O}_{1B}=\hat{b}_{1}$
(linear loss).

In order to further normalize the equation~(\ref{eq:w:master-eqn}),
we use the dimensionless time $\tau=g_{11}N_{A}t$ introduced in Section~\ref{sec:Dynamical-preparation-of}.
This gives us the dimensionless master equation
\begin{equation}
\frac{d\hat{\rho}}{d\tau}=-i[\tilde{H},\hat{\rho}]+\sum_{\bm{l}}\tilde{\gamma}_{\bm{l}}\mathcal{L}_{\bm{l}}[\hat{\rho}],\label{eq:w:master-eqn-dimensionless}
\end{equation}
with a dimensionless Hamiltonian $\tilde{H}$ as in Eq.~(\ref{eq:dimensionless-H}),
and dimensionless loss coefficients $\tilde{\gamma}_{\boldsymbol{l}}=\gamma_{\boldsymbol{l}}/g_{11}N_{A}$.
This equation can be transformed to the equivalent partial differential
equation by applying the Wigner transformation~\cite{wigner}:

\begin{eqnarray*}
\mathcal{W}\left[\hat{A}\right] & = & \frac{1}{\pi^{8}}\int d^{2}\lambda_{1}\, d^{2}\lambda_{2}\, d^{2}\lambda_{3}\, d^{2}\lambda_{4}\\
 &  & \quad\quad\times\left(\prod_{i=1}^{4}\exp\left(-\lambda_{i}z_{i}^{*}+\lambda_{i}^{*}z_{i}\right)\right)\\
 &  & \quad\quad\times\mathrm{Tr}\left\{ \hat{A}\prod_{i=1}^{4}\exp\left(\lambda_{i}\hat{z}_{i}^{\dagger}-\lambda_{i}^{*}\hat{z}_{i}\right)\right\} ,
\end{eqnarray*}
where 4-vectors $\boldsymbol{z}^{T}=\begin{pmatrix}\alpha_{1} & \beta_{1} & \alpha_{2} & \beta_{2}\end{pmatrix}$
and $\hat{\boldsymbol{z}}^{T}=\begin{pmatrix}\hat{a}_{1} & \hat{b}_{1} & \hat{a}_{2} & \hat{b}_{2}\end{pmatrix}$
were introduced for convenience. The resulting differential equation,
after truncating higher-order derivatives, is a Fokker-Planck equation
(FPE) for the truncated (positive) Wigner function $W\equiv\mathcal{W}\left[\hat{\rho}\right]$
\[
\frac{dW}{d\tau}=-\bm{\partial}_{\bm{z}}^{T}\boldsymbol{a}W-\bm{\partial}_{\bm{z}^{*}}^{T}\boldsymbol{a}^{*}W+\mathrm{Tr}\left\{ \bm{\partial}_{\bm{z}}\bm{\partial}_{\bm{z}^{*}}^{T}BB^{H}\right\} W,
\]
 where
\[
\boldsymbol{z}^{T}=\begin{pmatrix}\alpha_{1} & \beta_{1} & \alpha_{2} & \beta_{2}\end{pmatrix},
\]
\[
\boldsymbol{\partial}_{\bm{z}}^{T}=\begin{pmatrix}\frac{\partial}{\partial\alpha_{1}} & \frac{\partial}{\partial\beta_{1}} & \frac{\partial}{\partial\alpha_{2}} & \frac{\partial}{\partial\beta_{2}}\end{pmatrix},
\]
\[
\boldsymbol{a}=-i\boldsymbol{a}_{\mathrm{drift}}-\boldsymbol{a}_{\mathrm{loss}},
\]
\[
\bm{a}_{\mathrm{drift}}=\begin{pmatrix}\tilde{\kappa}_{1}\beta_{1}+\alpha_{1}\{\tilde{g}_{11}|\alpha_{1}|^{2}+\tilde{g}_{12}|\alpha_{2}|^{2}\}\\
\tilde{\kappa}_{1}\alpha_{1}+\beta_{1}\{\tilde{g}_{11}|\beta_{1}|^{2}+\tilde{g}_{12}|\beta_{2}|^{2}\}\\
\tilde{\kappa}_{2}\beta_{2}+\alpha_{2}\{\tilde{g}_{12}|\alpha_{1}|^{2}+\tilde{g}_{22}|\alpha_{2}|^{2}\}\\
\tilde{\kappa}_{2}\alpha_{2}+\beta_{2}\{\tilde{g}_{12}|\beta_{1}|^{2}+\tilde{g}_{22}|\beta_{2}|^{2}\}
\end{pmatrix},
\]
\[
\bm{a}_{\mathrm{loss}}=\begin{pmatrix}\alpha_{1}\{\tilde{\gamma}_{12}|\alpha_{2}|^{2}+\tilde{\gamma}_{1}\}\\
\beta_{1}\{\tilde{\gamma}_{12}|\beta_{2}|^{2}+\tilde{\gamma}_{1}\}\\
\alpha_{2}\{\tilde{\gamma}_{12}|\alpha_{1}|^{2}+2\tilde{\gamma}_{22}|\alpha_{2}|^{2}\}\\
\beta_{2}\{\tilde{\gamma}_{12}|\beta_{1}|^{2}+2\tilde{\gamma}_{22}|\beta_{2}|^{2}\}
\end{pmatrix},
\]

\[
B=\begin{pmatrix}\sqrt{\tilde{\gamma}_{12}}\alpha_{2} & 0 & 0 & 0 & \sqrt{\tilde{\gamma}_{1}} & 0\\
0 & \sqrt{\tilde{\gamma}_{12}}\beta_{2} & 0 & 0 & 0 & \sqrt{\tilde{\gamma}_{1}}\\
\sqrt{\tilde{\gamma}_{12}}\alpha_{1} & 0 & \sqrt{\tilde{\gamma}_{22}}\alpha_{2} & 0 & 0 & 0\\
0 & \sqrt{\tilde{\gamma}_{12}}\beta_{1} & 0 & \sqrt{\tilde{\gamma}_{22}}\beta_{2} & 0 & 0
\end{pmatrix}.
\]

This FPE is equivalent to the following set of stochastic differential
equations (SDEs)~\cite{risken_fpe}:
\begin{equation}
d\bm{z}=\bm{a}d\tau+Bd\bm{Z},\label{eq:w:sdes}
\end{equation}
where $d\bm{Z}^{T}=\begin{pmatrix}dZ_{12A} & dZ_{12B} & \cdots & dZ_{1A} & dZ_{1B}\end{pmatrix}$
is a complex 6-dimensional Wiener process. These equations can be
solved numerically using conventional methods, and their solution
can, in turn, be used to get the expectations of symmetrically ordered
operator products as
\begin{align*}
\langle\left\{ \left(\hat{a}_{i}^{\dagger}\right)^{m}\hat{a}_{j}^{n}\ldots\right\} _{\mathrm{sym}}\rangle & =\int\left(\alpha_{i}^{*}\right)^{m}\alpha_{j}^{n}\ldots W\, d^{2}\boldsymbol{z}\\
 & \approx\langle\left(\alpha_{i}^{*}\right)^{m}\alpha_{j}^{n}\ldots\rangle_{\mathrm{paths}},
\end{align*}
where $\langle\rangle_{\mathrm{paths}}$ stands for the average over
the simulation paths.

As for Section~\ref{sec:Dynamical-preparation-of}, we assume the
initial state to be the coherent state:
\begin{equation}
|{\psi}\rangle=|{\alpha_{0}}\rangle_{a_{1}}|{\alpha_{0}}\rangle_{a_{2}}|{\beta_{0}}\rangle_{b_{1}}|{\beta_{0}}\rangle_{b_{2}},\label{eq:w:coh}
\end{equation}
 where $\alpha_{0}=\sqrt{N_{A}/2}$, $\beta_{0}=\sqrt{N_{B}/2}$.
Therefore
\[
\alpha_{i}=\alpha_{0}+\frac{1}{2}\eta_{1i},\,\beta_{i}=\beta_{0}+\frac{1}{2}\eta_{2i},
\]
 where $\eta_{1i}$ and $\eta_{2i}$ are complex normally distributed
random numbers with $\langle\eta_{ji}^{*}\eta_{kl}\rangle=\delta_{jk}\delta_{il}$.

\end{document}